\documentclass[twocolumn]{aastex6}
\usepackage{natbib}
\usepackage{color}
\usepackage{soul}
\usepackage{graphicx}
\usepackage{url}
\usepackage{amsmath}
\usepackage{xcolor}
\usepackage{ulem}

\begin{document}
\title{Star formation and evolution of blister-type H{\sc ii} region Sh2-112}
\author{Neelam Panwar \altaffilmark{1}, Saurabh Sharma \altaffilmark{1}, D. K. Ojha \altaffilmark{2}, Tapas Baug \altaffilmark{3}, L. K. Dewangan \altaffilmark{4}, B. C. Bhatt \altaffilmark{5}, Rakesh Pandey \altaffilmark{1}}
\altaffiltext{1}{Aryabhatta Research Institute of Observational Sciences (ARIES), Nainital - 263129, India}
\altaffiltext{2}{Tata Institute of Fundamental Research, Mumbai (Bombay) - 400005, India}
\altaffiltext{3}{Kavli Institute for Astronomy and Astrophysics, Peking University, 5 Yiheyuan Road, Haidian District, Beijing 100871, People's Republic of China}
\altaffiltext{4}{Physical Research Laboratory, Navrangpura, Ahmedabad 380009, India} 
\altaffiltext{5}{Indian Institute of Astrophysics, Koramangala, Bangalore 560034, India}

\begin{abstract}
 	We report the observational findings of the Sh2-112 H{\sc ii} region by using the 
	multiwavelength  data analysis ranging from optical to radio wavelengths. 
	This region is powered by a massive O8V-type star BD +45 3216.  
	The surface density distribution and minimum spanning tree analyses of the young stellar object (YSO) candidates in the region reveal their 
	groupings toward the western periphery of the H{\sc ii} region. 
	A GMRT radio continuum emission peak is found toward the north-west boundary of the H{\sc ii} region and
	is investigated as a compact/ultra-compact H{\sc ii} region candidate powered by a B0-B0.5 type star. 
	Toward the south-west direction, a prominent curved rim-like structure is found in the H$\alpha$ image 
	and GMRT radio continuum maps, where the H$_2$ and $^{13}$CO emission is also observed. These results 
	suggest the existence of the ionized boundary layer (IBL) on the surface of 
	the molecular cloud. This IBL is found to be over-pressurized with respect to the internal pressure of the surrounding molecular cloud. 
	This implies that the shocks are propagating/ propagated into the molecular cloud and the young stars
identified within it are likely triggered due to the massive
star. 
	It is also found that this region is 
	ionization bounded toward the west-direction and density bounded toward the east-direction. 
	Based on the distribution of the ionized gas, molecular material, and the YSO candidates; we propose that the Sh2-112 H{\sc ii} region is a good candidate for the blister-type H{\sc ii} region which has been 
	evolved on the surface of a cylindrical molecular cloud. 
\end{abstract}

\section{Introduction}
 Massive stars play dominant roles in the dynamical and the chemical evolution of the host galaxies. 
 Even in their short life-span, they influence their surrounding environment through large output 
 of ultraviolet (UV) photons, strong stellar winds, outflows and energetic supernova
 explosions \citep{zinn07,krui19}. 
 Massive stars create H{\sc ii} regions by ionizing the surrounding gas through their energetic 
 radiation ($h\nu$ $\ge$ 13.6 eV). 
 The interaction of the ionization/shock fronts from the expanding H{\sc ii} regions with the surrounding molecular material may trigger the formation 
 of next generation stars via various processes \citep[see, e.g.,][]{elme98,deha05}. 
In one of these processes, known as `collect \& collapse', a dense shell of swept up neutral material forms around the massive stars which becomes gravitationally unstable and fragments to form new stars 
\citep{elme77,whit94}. The outcome of the above process is observed as the massive condensations or groups of young stars at the periphery \citep[e.g.,][]{deha05,zava10}. 
In another process known as `radiation driven implosion (RDI)', ionization/shock fronts can compress the 
pre-existing dense clumps and consequently induce the formation of new stars \citep{bert89,lefl94}. 
The surface of the dense clump may shield the remaining molecular cloud from 
the ionizing radiation, resulting in the structures, such as, bright-rim clouds, globules, pillars, etc. 

Though, there are observational studies of a few star-forming regions (SFRs) investigating the influence
of massive stars on their surroundings \citep[see][and references therein]{zava10,poma09,baug15,panwar2017,sharma17,dewangan19,pandey20}, 
however, the physical processes involved in the formation of massive stars and their interaction with 
surrounding environment are observationally not very well established \citep{tan14,mott18}. The numerical simulations show that an 
expanding H{\sc ii} region will be able to trigger star formation if the ambient
molecular material is dense enough \citep{hoso06,dale07}. 
However, observationally it is not well studied, especially in the case of clumpy and inhomogeneous environments. 

Sh2-112 \citep{sharpless59} is a Galactic H{\sc ii} region ($\alpha$$_{2000}$ 
$\sim$ 20h33m49s, $\delta$$_{2000}$ $\sim$ +45$^\circ$38$\arcmin$00$\arcsec$) 
powered by a massive star BD +45 3216 \citep{lahu85,hunt90}. 
It is located towards one of the most active and massive star-forming complexes within 2 kpc distance in the Galaxy; 
the Cygnus X, which is extended in an area of $\sim$ 7$^\circ$ $\times$ 7$^\circ$ on the sky. 
Although once believed to be a superposition of many disconnected SFRs, \citet{schn06} showed that the molecular clouds in Cygnus~X 
form a coherent complex of 9 OB associations at the similar distance of $\sim$ 1.7 kpc. 
The entire complex exhibits evidence for many sites of star formation at
 different evolutionary stages, from the youngest embedded star formation in an infra-red (IR) dark clouds 
 in DR21 \citep{down66} to the young cluster Cygnus OB2 \citep{knod00} to the more dispersed 
and perhaps older Cygnus OB9 region. Sh2-112 is associated with the Cygnus OB6, adjacent to the North American and 
Pelican Nebula, but at farther distance \citep{uyan01}. Among all the regions of the Cygnus X, due to a complex morphology, Sh2-112 is a poorly studied 
H{\sc ii} region till now. 

\begin{figure*}
\centering
\includegraphics[width=0.7\textwidth]{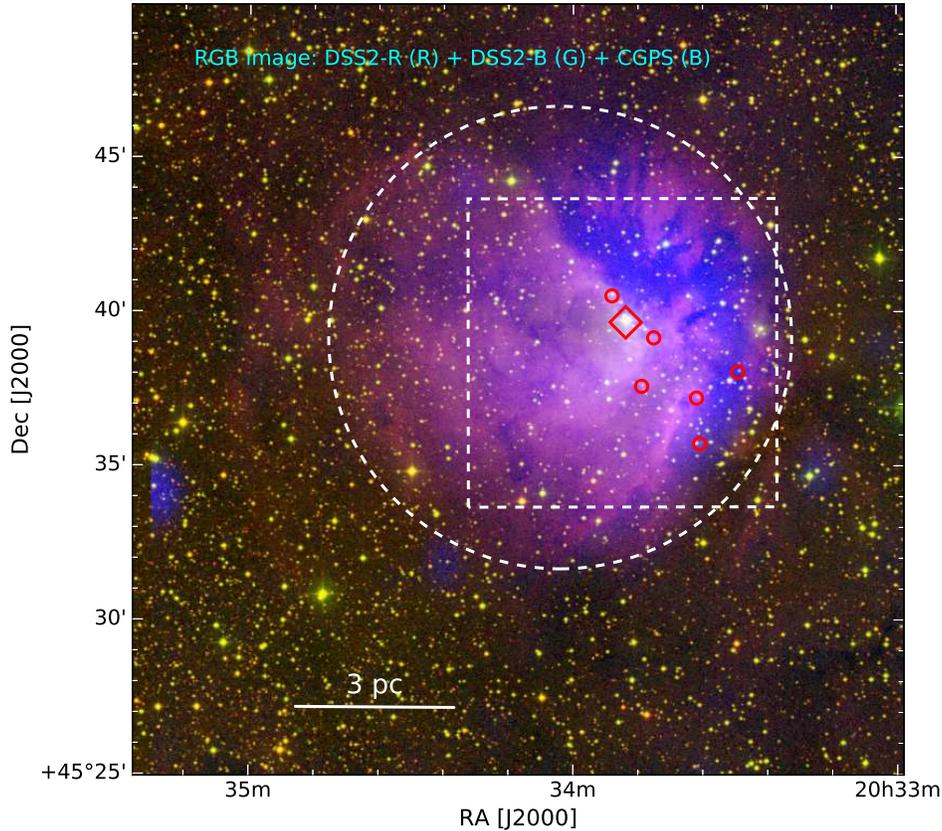}
	\caption{Color-composite image of the Sh2-112 region generated from the
DSS2-R (red), DSS2-B (green), CGPS 21 cm radio continuum (blue) images. 
The diamond represents the position of the ionizing star. Dashed circle shows the extension of the H{\sc ii} region (diameter $\sim$ 15$\arcmin$) in optical images. Small red circles represent the H$\alpha$ emission stars identified using slitless spectroscopy within the HCT FOV shown by dashed box (10$^\prime$ $\times$ 10$^\prime$).}
\label{fig1}
\end{figure*}

Fig. \ref{fig1} shows the color-composite view of the Sh2-112 region and
reveals that the complex has a spherical shell-like morphology with a diameter of $\sim$ 15$^\prime$. 
In addition to the observed interesting morphology, the complex is a 
relatively nearby star-forming site, making it a promising region to study the role of radiation feedback from massive star(s) in the evolution of natal cloud. 
The distribution of the young stellar object (YSO) candidates helps us study the current sites of star formation 
and relationships between high-mass stars and star formation activity in their parental molecular clouds. Determining how many low-mass stars form in clusters or in relative isolation will improve our understanding on how low-mass stars form in molecular cloud complexes dominated by massive stars.

In the present work, we aim to identify and characterize the YSO candidates in the H{\sc ii} 
region and study the star formation history using multiwavelength data covering from optical to radio wavelengths. 
Such analysis offers an opportunity to examine the distribution of YSO candidates, dust temperature, column density, extinction, ionized emission, and molecular gas.

This paper is organized as follows. In Section 2, we present the multiwavelength observations of the Sh2-112 region 
and data reduction techniques. Other available archival data sets used in the present work
 are also summarized in Section 2. In Section 3, we present the optical spectrum of the brightest star and identification/characterisation of YSO candidates. Morphology of the region inferred using multiwavelength observations
 followed by a discussion on star formation scenario is presented in Section 4. 
 In Section 5, the main conclusions of the present study are summarized.

\section{Observations and Data Reduction}
\label{sec:reduc}
To understand the ongoing star formation processes in the Sh2-112 complex, 
we have utilized the available data starting from the optical to radio wavelengths. The details 
of these data are briefly described in the following subsections.
\subsection{Optical observations}
Optical [S {\sc ii}] (6724 \AA), H$\alpha$ (6563 \AA),
and [O {\sc iii}] (5007 \AA) narrow-band filter observations of the central region of the Sh2-112 H{\sc ii} region were 
carried out on 2010 October 12 using the Himalaya Faint Object 
Spectrograph and Camera (HFOSC) mounted on the 2-m Himalayan $Chandra$ 
Telescope (HCT) \citep[for details, see,][]{stalin08,chauhan09}. With an integration time of
300s in each filter, the HCT observations cover a field of view (FOV) $\sim$ 10$\arcmin$ $\times$ 10$\arcmin$. 
Along with the object frames, several bias frames and twilight flat-field frames were also obtained. The average {\sl seeing} during the observations was 
$\sim$ 1$\arcsec$ - 1$\arcsec$.5. 

We carried out the short-exposure optical ($UBVI$) observations of the Sh2-112 region 
(FOV $\sim$ 18$\arcmin$ $\times$ 18$\arcmin$) on 
2018 October 05 using
2k$\times$2k CCD camera mounted on f/4 Cassegrain focus of the
1.3-m Devasthal Fast Optical Telescope (DFOT) of Aryabhatta Research
Institute of Observational Sciences (ARIES), Nainital, India \citep[for details, see][]{panwar2017}.
The average seeing during the observing night was $\sim2^{\prime\prime}$.
The log of observations is given in Table 1.
Along with the object frames, several bias and flat frames were also taken
during the same night. We also observed Landolt standard stars' field (SA 92) on the same night. The standard field was used to calculate
the extinction coefficients and calibrate our observations to the standard system.

Initial processing of the data frames (i.e., bias subtraction, flat fielding, etc.) was done using the IRAF\footnote{IRAF is distributed by National Optical Astronomy
Observatories, USA} data reduction package. The photometric measurements of the stars were performed using $DAOPHOT-II$
software package (Stetson 1987). The point spread function (PSF)
was obtained for each frame using several uncontaminated stars.
Aperture photometry was carried out for the standard stars to estimate the atmospheric extinction and to calibrate the observations. We found the value of extinction coefficients in $U$, $B$, $V$ and $I$ filters as 0.50, 0.26, 0.17 and 0.09, respectively. The following transformation equations were used to calibrate the
observations:\\
	($U$ - $B$) = (1.00$\pm$0.02)($u$ - $b$) + (-1.54$\pm$0.02)\\
	(B - V) = (1.22$\pm$0.02) (b - v) + (-0.86$\pm$0.02)\\
	(V - I) = (0.94$\pm$0.02) (v - i) + (0.20$\pm$0.01)\\
	B = b + (0.07$\pm$0.02) (u - b) + (-2.90$\pm$0.02)\\
	V = v + (-0.11$\pm$0.02) (v - i) + (-2.22$\pm$0.02)\\

where $u$, $b$, $v$, $i$ are the instrumental magnitudes corrected for the atmospheric extinctions, and $U$, $B$, $V$, $I$ are the standard magnitudes, respectively.

The standard deviations of the standardization residual, $\Delta$, between standard and transformed $V$ magnitude
and $(U-B)$, $(B-V)$ and $(V-I)$ colors of standard stars are 0.01, 0.02, 0.02 and 0.02 mag, respectively. 
We have used only those stars for further analyses which have photometric uncertainty $<$0.1 mag. 

\begin{table}
\caption{Log of optical observations}
\begin{tabular}{|p{1.8in}|p{1.4in}|}
\hline
	Filter  \& Exposure(sec)$\times$no. of frames& Date of observations (Telescope)\\
\hline
	[S {\sc ii}]: 300$\times$1, H$\alpha$: 300$\times$1, and [O {\sc iii}]: 300$\times$1&  2010 October 12  (HCT)\\
	U: 15$\times$3, B: 5$\times$2, V: 5$\times$3, I: 1$\times$2 & 2018 October 05 (DFOT)\\
	{\bf Slitless spectra} &\\
	H$\alpha$/gr 5: 300$\times$2&  2010 October 12 (HCT)\\
	{\bf slit spectra} &\\
	BD +45 3216: 600$\times$2&  2010 October 12 (HCT)\\
	Fiege 15: 600$\times$2&  2010 October 12 (HCT)\\
\hline
\end{tabular}
\label{tab1}
\end{table}
\subsection{Optical slitless spectroscopy}
We performed H$\alpha$ slitless spectroscopic observations of the Sh2-112 
using the HFOSC on the 2-m HCT on 2010 October 12. The spectra were
obtained by the combination of a wide H$\alpha$ filter (6300-6740 
\AA) and grism 5 (5200-10300 \AA) with a spectral resolution of
870 \AA. The average seeing during the observations was $\sim$ 1$\arcsec$.2. Two frames of slitless spectra,
each with an exposure time of 300 seconds, were obtained and 
co-added to increase the signal-to-noise (S/N) of the observed spectra.
Emission line stars were visually identified as an enhancement
of H$\alpha$ emission above the continuum. The six sources identified as emission-line stars within the HCT FOV are shown with red circles in Fig. \ref{fig1}.

\subsection{Optical Slit Spectroscopy}
To ascertain the spectral type of the brightest star, BD +45 3216 ($\alpha$$_{2000}$: 20h33m50.4s, 
$\delta$$_{2000}$: 45$^\circ$39$\arcmin$41$\arcsec$) in the field, we 
obtained the spectrum using
 HFOSC with the help of Grism 7 (3500 to 7000 \AA), which
has a resolving power of 1200 and a spectral dispersion of
1.45 \AA/pixel. The spectrum of BD +45 3216 with an exposure time of 600s was taken on 2010 October 12. 
In addition to FeAr lamp arc spectrum (for wavelength calibration), multiple bias frames were also
obtained. The spectro-photometric standard star Feige 15 \citep[$\alpha_{2000}$: 01h49m09.4s, $\delta_{2000}$: +13$^\circ$33$\arcmin$12$\arcsec$;][]{ston77} was also observed 
with an exposure time of 600s. We reduced all the spectra using `APALL' task in IRAF data reduction package. Finally, the flux-calibrated normalized spectrum of the bright star was obtained (see Figure \ref{fig2}).

\subsection{Radio Continuum Observations}
Radio continuum observations of the Sh2-112 region at 610 and 1280 MHz bands were carried out using 
the Giant Metrewave Radio Telescope (GMRT) on 2010 October 22 (Project Code 19${\_026}$; P.I.- S. S. Borgaonkar) and 2012 
November 09 (Project Code 23${\_019}$; P.I.- K. K. Mallick), respectively. 
Total observing time at 610 and 1280 MHz bands was 5.4
hours and 4.5 hours, respectively. The Astronomical Image Processing Software (AIPS)
package was used following the procedure described in \citet{mall13}.
Various AIPS tasks were used to edit the data and to flag out the bad baselines or bad
time ranges. Additionally, the data quality was improved by multiple iterations of flagging
and calibration. A few iterations of
(phase) self-calibration were carried out to remove the ionospheric
phase distortion effects. Finally, the data were Fourier-inverted to make the radio maps. The final 610 MHz image has 
a synthesized beamsize 60$\arcsec$ $\times$ 60$\arcsec$ and rms of 48 mJy/beam. The final rms of 1280 MHz image is about 0.13 mJy/beam and the beam size is 2$\arcsec$.5 $\times$ 2$\arcsec$.2 , respectively. 
Both 610 and 1280 MHz images were corrected for the system temperature  
\citep[see][]{omar02,mall12,vig14} and 
both the images were rescaled by a correction factor \citep[see][]{baug15, dewa18}.
\subsection{Archival Data}
\subsubsection{Near-infrared JHK data from UKIDSS and 2MASS}
United Kingdom Infrared Deep Sky Survey (UKIDSS) archival data of the
 Galactic Plane Survey \citep[GPS release 6.0;][]{lawr07} are available for the Sh2-112
 region. UKIDSS observations were obtained using the UKIRT Wide Field Camera (WFCAM). 
Following the selection procedure of the GPS photometry discussed in
\citet{dewa15}, we retrieved only reliable near-infrared (NIR) sources (photometric uncertainty $<$ 0.1 mag) in the region. We have also used the NIR data 
from Two-Micron All Sky Survey \citep[2MASS;][]{cutr03}. 
To calibrate the UKIDSS $J$, $H$, $K$ photometric system to the 2MASS system, 
we rescaled all UKIDSS magnitudes according to their ($J$ - $H$) and ($H$ - $K$) UKIDSS colors. 
We merged the 2MASS and UKIDSS catalogs into a single NIR
source catalog, choosing the NIR magnitudes of the fainter sources from the more precise UKIDSS 
magnitudes.
In general, bright sources are saturated in the UKIDSS frames, and thus in the final catalog, magnitudes for the sources brighter than J $\sim$ 14 mag are replaced by the 2MASS magnitudes.
 For further analysis, we considered sources with good photometric magnitudes, i.e., having accuracy better than 10\%.

\subsubsection{Wide Infrared Survey Explorer data}
The Wide-field Infrared Survey Explorer (WISE) has mapped the sky in four wavebands (3.4, 4.6, 12, 
and 22 $\micron$) and uncovered populations of YSO candidates hindered in the dense clouds. 
We used the $WISE$ catalog from \citet{cutr14} to identify YSO candidates in the Sh2-112 region. The spatial resolution at first three WISE bands is 6$^{\prime\prime}$ 
 and reaches 12$^{\prime\prime}$ at 22 $\mu$m.  
To ensure the good quality photometry, we considered only those sources which have magnitude uncertainties $\le$ 0.2 mag and rejected sources with contamination and 
confusion flags (cc-flags in the catalog) that include 
any of ``D'', ``H'', ``O'' or ``P''. The 3.4 $\micron$ and 12 $\micron$ bands include prominent polycyclic aromatic hydrocarbon (PAH) 
features at 3.3, 11.3, and 12.7 $\micron$ \citep{sama07,wrig10} in addition to the continuum emission, and hence can be used to get an idea of the photo-dissociation region (PDR). The 22 $\micron$ band can be used to examine the warm dust emission, i.e., the stochastic emission from small grains as well as the thermal emission from large grains \citep{wrig10}.  The WISE catalog also includes corresponding 2MASS magnitudes in the $JHK_s$ bands of the sources detected in WISE wavebands. 

\subsubsection{21 cm radio continuum data from CGPS} 
We retrieved 21 cm radio continuum data from the Canadian Galactic Plane Survey \citep[CGPS;][]{taylor03} to trace the extension of ionized gas around Sh2-112.  
\subsubsection{$^{13}$CO(J =3 - 2) data}
$^{13}$CO (J = 3 - 2) (330.588 GHz; beam $\sim$ 14$^{\prime\prime}$) data were retrieved from the 
James Clerk Maxwell Telescope (JCMT) archive (ID: M08AU19; PI: Stuart Lumsden). The observations were obtained
 in position-switched raster-scan mode of the Heterodyne Array Receiver Program
 \citep[HARP;][]{buck09}. We utilized the processed integrated $^{13}$CO intensity map of the western part of the Sh2-112 region.
\subsubsection{Other Datasets}
{\bf Optical and Near-infrared images:}\\
We utilized the wide-field H$\alpha$-, I- and B- band images from the National Optical Astronomical Observatory (NOAO) archive to trace the extension of the ionized hydrogen and optical morphology of 
the region.

We also accessed the processed H$_2$- and K-band images from the Canada-France-Hawaii Telescope archive observed with the Cam\'era Panoramique Proche InfraRouge \citep[CPAPIR; see][]{arti04} imager attached to the Observatoire du Mont-M\'egantic 1.6-m telescope operated by the Universit\'e de Montre\'al, Université Laval. CPAPIR is based on a 2048 $\times$ 2048 pixel$^2$ Hawaii-2 IR 
array detector. With a pixel size of 0$\arcsec$.89, it has a field of 
view of $\sim$ 30 $\times$ 30 arcmin$^2$. 

{\bf GAIA DR2 data :} We utilized the Gaia Data Release 2 \citep{gaia18} from the 
European Space Agency (ESA) mission Gaia to estimate the proper motion of the stars and distance of the complex.
\section{Results}
\subsection{Spectral Analysis of the Brightest Source}
\begin{figure}
\centering
\includegraphics[scale = 0.48, trim = 20 0 0 160, clip]{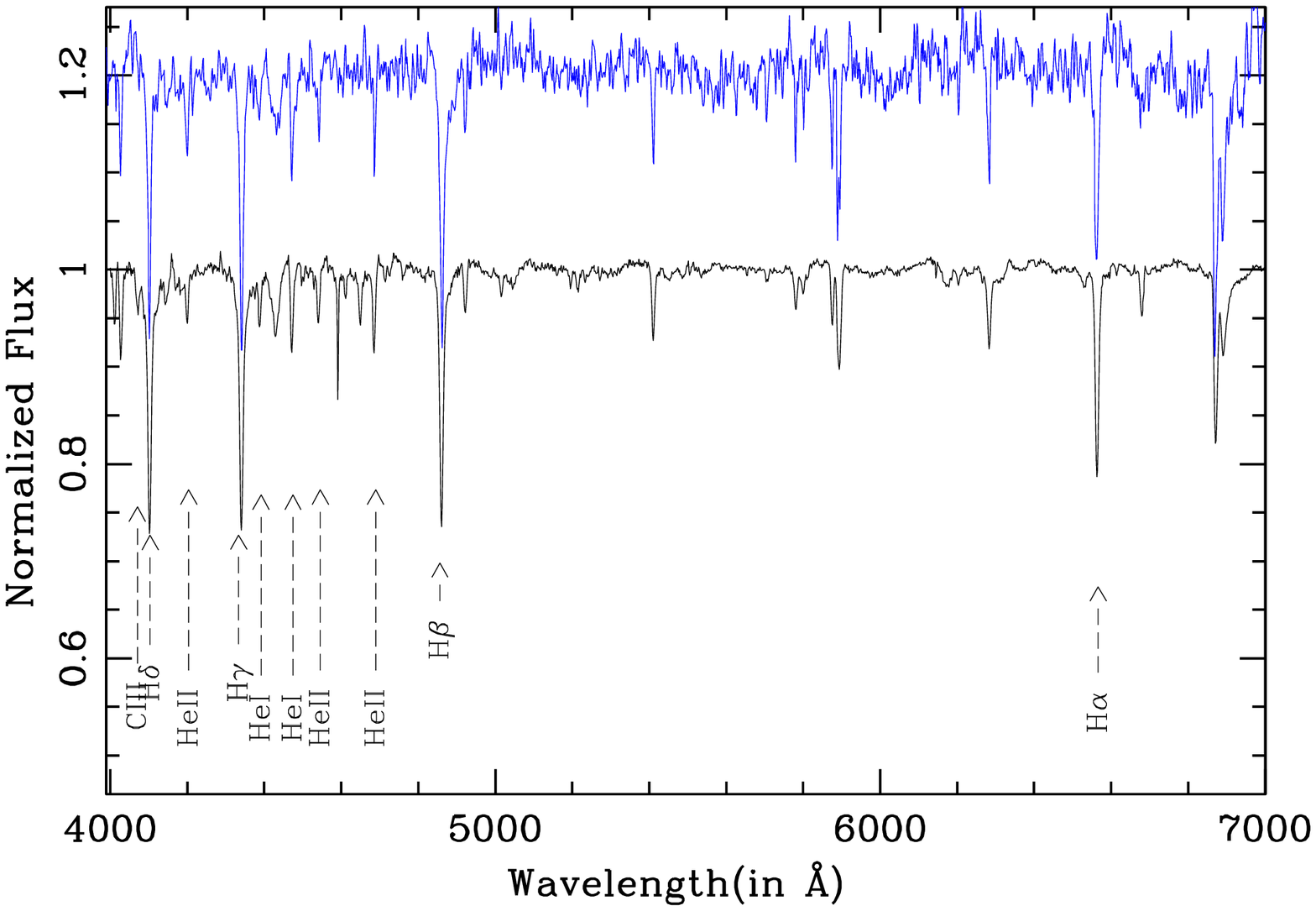}
	\caption{Flux-calibrated normalized spectrum for the optically bright source in the Sh2-112. Important spectral lines are marked in the figure. The spectrum of an O8V star (blue curve) from \citet{jaco84} is also shown for comparison.}
	\label{fig2}
\end{figure}

In order to confirm the spectral type of the brightest source, BD +45 3216, we extracted the low-resolution, 
one-dimensional spectrum of the source. The flux-calibrated, normalized spectrum of the brightest source 
in the wavelength range of 4000 - 7000 \AA~is shown in Fig. \ref{fig2}. 
For spectral classification of the star, we used the criteria
given by \citet{walb90}. Spectra of O and B stars have the features
of hydrogen, helium and other atomic lines (e.g. O {\sc ii}, C {\sc iii}, Si {\sc iii}, Si {\sc iv},
Mg {\sc ii}). Hydrogen and helium lines are usually seen in absorption for
dwarfs whereas they may be in emission in supergiants. For the spectral classification 
of the star, we use hydrogen, He {\sc ii} and He {\sc i} lines. The presence of He {\sc ii} lines 
(4686 \AA, 5411 \AA) and He {\sc ii}+{\sc i} (4026 \AA) limits
the spectral type to O type. 
Moderate nitrogen enhancement
indicates a later O-type spectrum. The line ratios of He {\sc ii} 4686 \AA/
He~{\sc i} 4713 \AA, Si {\sc iv} 4089 \AA/He {\sc i} 4144 \AA, 4387 \AA, 4471 \AA, 4713 \AA, and
Si {\sc iv} 4116 \AA/He {\sc i} 4144 {\AA} suggest the star may be of O8 - O9V type. In the case
of early-type stars, the ratio of He {\sc i} 4471 \AA/He {\sc ii} 4542 {\AA}  is a primary
indicator of the spectral type and the ratio is greater than 1 for
spectral type later than O7. The line strength of He~{\sc ii} gets weaker
for late O-type stars and He {\sc ii} (4686 \AA) is last seen in B0.5-type stars
\citep{walb90}. 
Finally, the
spectral type O8V was assigned to the source by a visual comparison
to the standard library spectra \citep{jaco84}. For comparison, we have also shown the spectrum of an O8V star (blue curve) from \citet{jaco84} in Fig. \ref{fig2}. 
Here, we note that an uncertainty of $\pm$ 1 in the sub-class identification is expected as the present analysis is based on the low-resolution spectrum of the star.
\begin{figure}
\centering
	\includegraphics[scale = 0.45, trim = 0 0 0 0, clip]{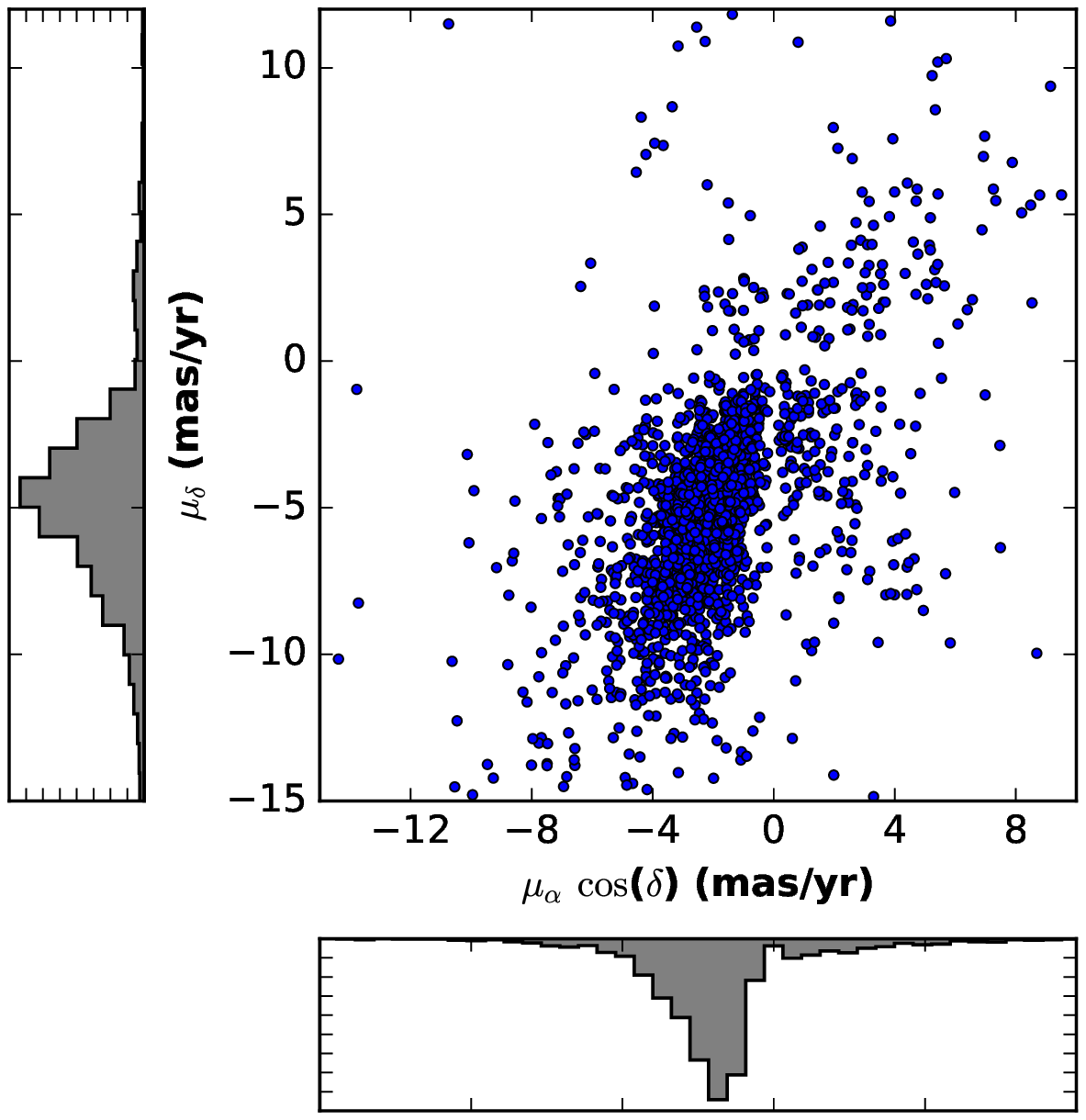}
\includegraphics[scale = 0.45, trim = 0 0 0 0, clip]{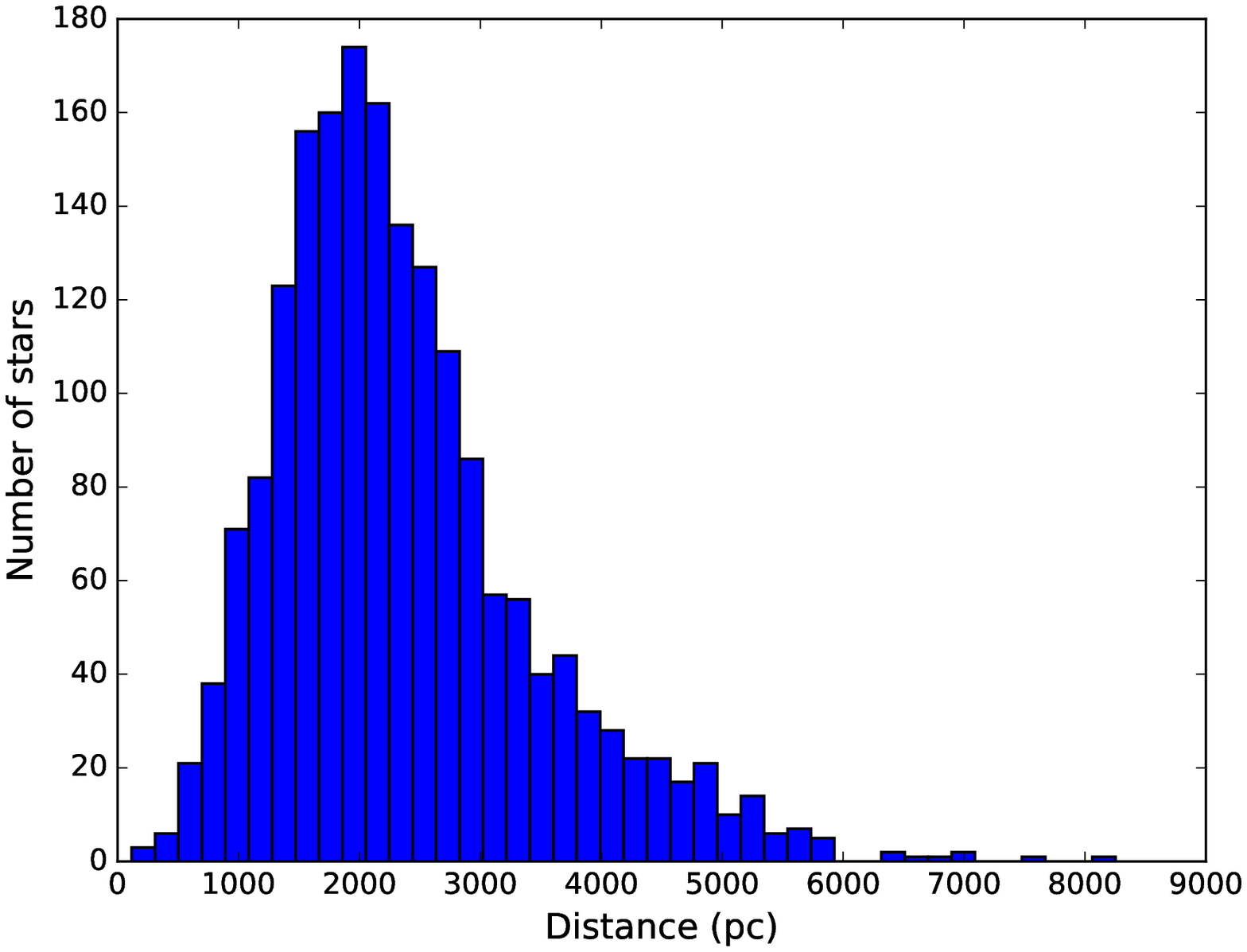}
	\caption{Top panel: Histograms of the proper motions of stars in the Sh2-112, taken from the Gaia DR2. 
	Bottom panel: Histogram of the distances of the stars having proper motion values within 3 $\sigma$ of the mean proper motions taken from \citet{jone18}.} 
	\label{fig4a}
\end{figure}
\subsection{Distance and reddening towards the region}
\begin{figure}
\centering
	\includegraphics[scale = 0.53, trim = 30 0 0 10, clip]{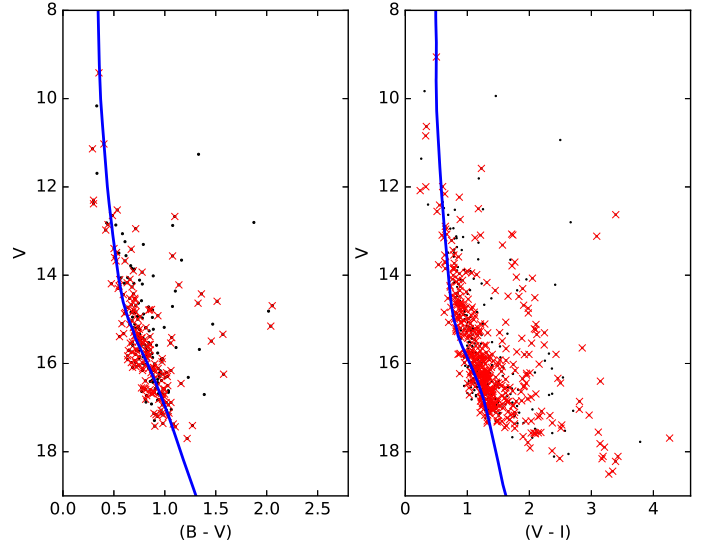}
	\caption{ ($B$ - $V$)/$V$ (left panel) and ($V$ - $I$)/$V$ (right panel) CM diagrams of the sources detected in the region using DFOT observations. The blue solid curves represent the zero-age main sequence (ZAMS) from \citet{girardi02}, corrected for 
a distance of  $\sim$ 2 kpc and reddening E($B$ - $V$) of $\sim$ 0.64 mag. Red crosses represent the sources with proper motions within 3 $\sigma$ of 
	the mean values (see the text).} 
	\label{optcmd}
\end{figure}
We use proper motions of the sources in the Sh2-112 region from the Gaia DR2 database \citep{gaia18} to identify 
the probable members of the region and to further determine the distance to the region. 
We used bright stars ($G$ $<$ 19 mag) within $\sim$ 27$\arcmin$ $\times$ 27$\arcmin$ and having 
good proper motion information ($>$ 5 $\sigma$) to construct the 
$\mu$$_\delta$ vs $\mu$$_\alpha$$\cos$($\delta$) vector point diagram (VPD). In Fig. \ref{fig4a} (top panel) we have shown the VPD along with the histograms of proper motions of the stars in the region. 
The Gaussian distribution fit to the proper motions along the right ascension ($\mu$$_\alpha$ cos($\delta$)) 
and declination ($\mu$$_\delta$) resulted in a mean value of -1.62 and -4.78 mas yr$^{-1}$ 
with a half-width at half maximum of 1.09 and 2.44 mas yr$^{-1}$, respectively. 
\citet{luri18} suggested that systematics and correlations in
the Gaia astrometric solution tend to overestimate the true distance and hence a Bayesian approach should be used 
to properly account for the covariance 
uncertainties in the parallaxes and proper motions from the Gaia DR2. \citet{jone18} have determined 
the distances for the stars using a Bayesian inference method. 
We used the \citet{jone18} catalog to obtain the distance of the H{\sc ii} region using distances of all the stars 
 located within our target area having proper motions within 3$\sigma$ of the mean values (see Fig. \ref{fig4a}, bottom panel). The mean distance of the H{\sc ii} region is estimated as $\sim$ 2.0 $\pm$ 0.7 kpc. The proper motions of the massive star 
 ($\mu$$_\alpha$ cos($\delta$) $\sim$ -4.19 $\pm$ 0.66 mas ${yr}^{-1}$ and $\mu$$_\delta$ $\sim$ -7.227 $\pm$ 0.66 mas ${yr}^{-1}$) are within 3$\sigma$ 
 of the mean values.  

We have also calculated the spectro-photometric distance to
the bright star BD +45 3216. The optical and the NIR magnitudes of the source are obtained from the DFOT and the 2MASS data, respectively. To estimate the
distance to the O8V source, we first obtained the intrinsic color
and magnitude of an O8V star (($B$ - $V$)$_0$ $\sim$ -0.26 and M$_V$ = -4.4 mag) from \citet{martins05}. 
We considered the mean R$_V$ of 3.1 for the reddening correction. 
Accordingly, the E ($B$ - $V$) of the source was found to be $\sim$ 0.64 mag. 
With an absolute magnitude (M$_V$) of -4.4 and
apparent magnitude ($V$) of 9.06, we estimated the distance to
the O8V star of $\sim$ 2 kpc. 
Note that large errors (at least 20\%) in these distance estimates could be 
due to photometric uncertainties, and the general extinction law used in the
estimation. The spectro-photometric distance is comparable to the estimated distance of the massive star ($\sim$ $1.8^{+1.7}_{-0.7}$ kpc) derived from the Gaia DR2 data \citep{jone18}.

In Fig. \ref{optcmd}, we have shown the optical color-magnitude (CM) diagrams of the sources detected in the region using the DFOT observations. The sources having proper motions within the 3$\sigma$ of mean proper motion 
values are shown with cross symbols. 
The blue solid curve represents the zero-age main sequence (ZAMS) from \citet{girardi02}, corrected for 
a distance of  $\sim$ 2.0 kpc and reddening E($B$ - $V$) of $\sim$ 0.64 mag, fits well to the CM diagrams. We notice a broad distribution of sources in the CM diagrams which may be due to variable 
reddening, young stages of stars, binaries or field star contamination etc. However, it is 
difficult to bifurcate the members of the H{\sc ii} region and field stars based on optical observations only. In the absence of spectroscopic observations, the robust approach to select the sources associated with 
the region is to identify young stellar content using multiwavelength photometric observations. 

\subsection{Young Stellar Population in the Region}
The young stellar population in a star-forming region is also helpful to trace the star formation 
processes in the region. In the present study, we have used the IR data from the WISE, 2MASS and UKIDSS to identify and characterize the young stars in the region. The various schemes are described below.
\subsubsection{WISE color-color space}
\begin{figure*}
\centering
\includegraphics[scale = 0.4, trim = 0 0 0 0, clip]{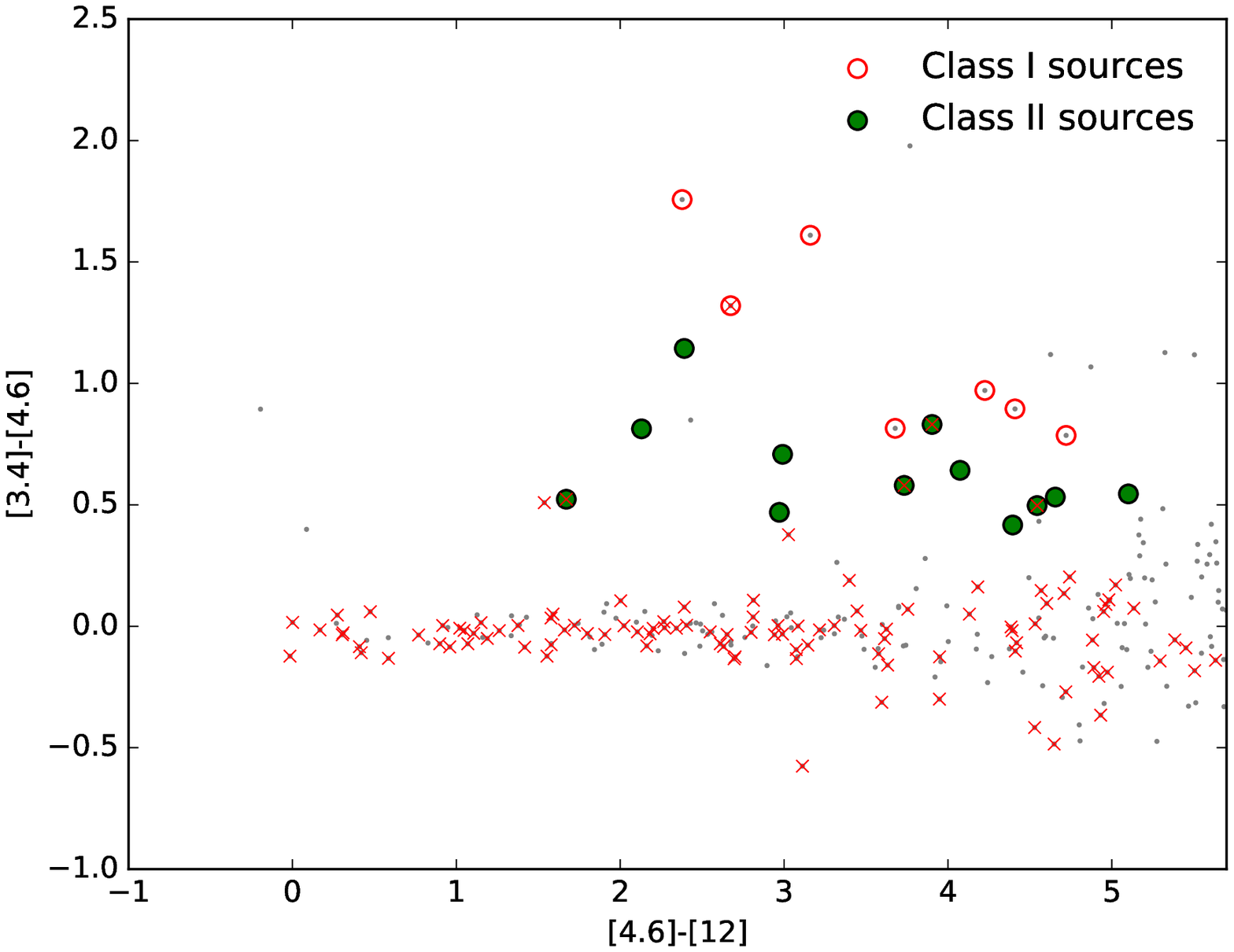}
\includegraphics[scale = 0.4, trim = 0 0 0 0, clip]{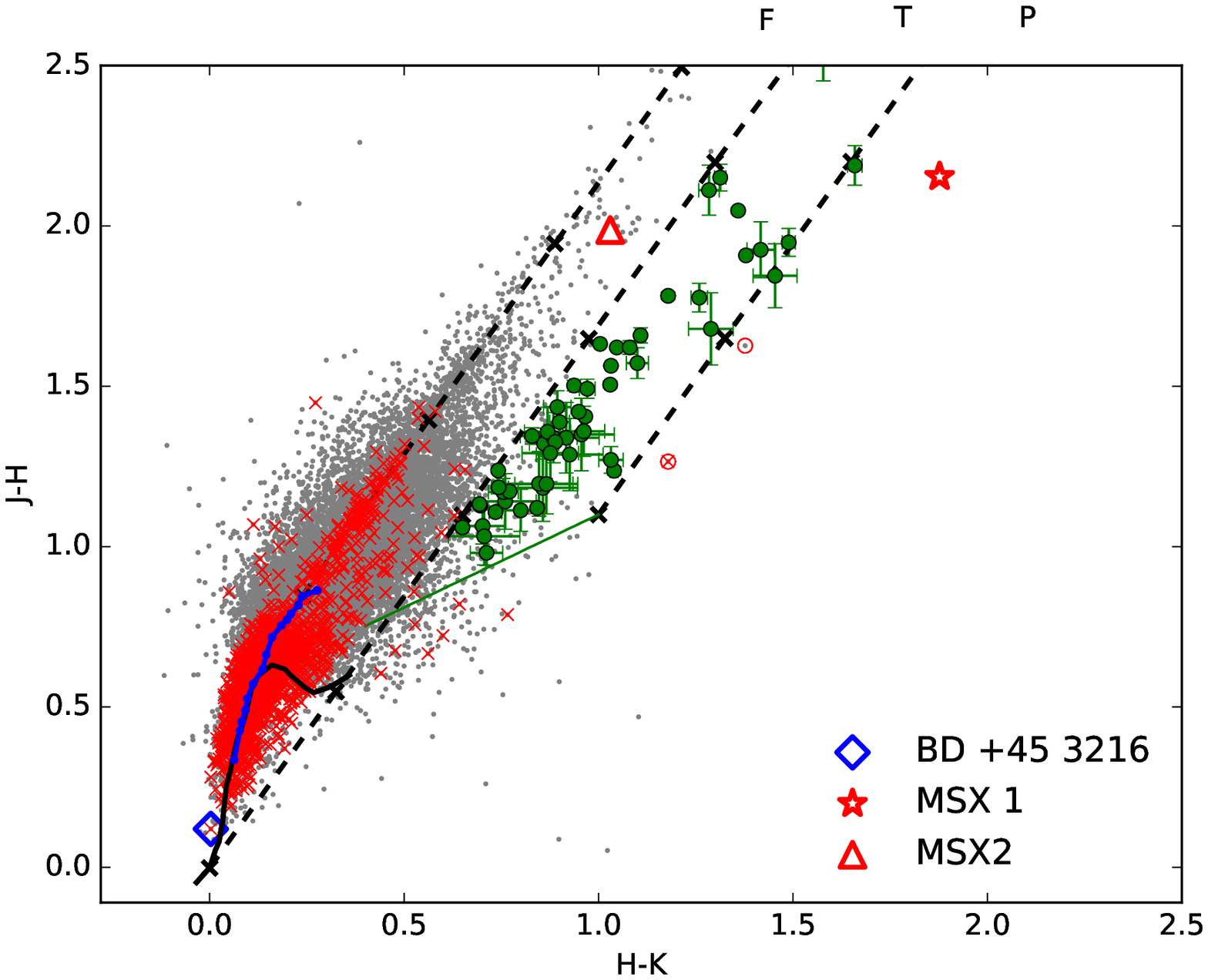}
\includegraphics[scale = 0.4, trim = 0 0 0 0, clip]{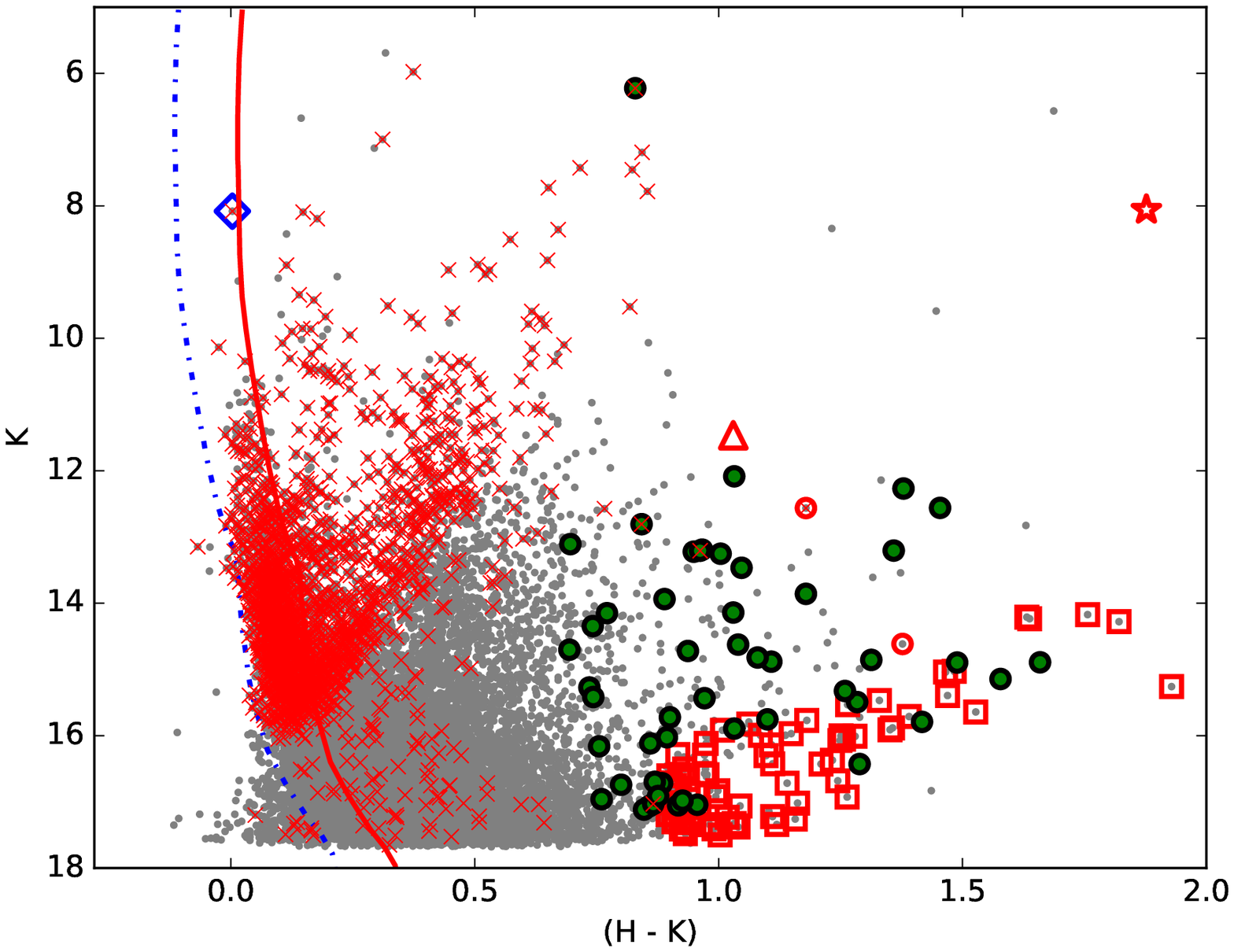}
	\caption{Top-left panel: [3.4]-[4.6]/[4.6]-[12] CC diagram for the sources in the Sh2-112 H{\sc ii} region. Filled and open circles represent the Class~{\sc ii} and Class~{\sc i} sources, respectively.
	Top-right panel : ($J$-$H$)/($H$-$K$) CC diagram for the sources in H{\sc ii} region. The thin black curve in the CC diagram represents the main-sequence (MS) locus and the thick blue curve is the locus of the giant \citep{bessel88}. The thin green line is the intrinsic locus of CTTS \citep{meyer97} and parallel dashed lines are the reddening vectors drawn from the 
	base of the MS locus, turning point of the MS locus, and the tip of the CTTS locus, respectively. 
	Bottom panel : $K$/($H$-$K$) CM diagram for the sources in the H{\sc ii} region. In the NIR CC and CM diagrams, filled and open circles are 
	Class {\sc ii} and Class {\sc i} (NIR excess) sources selected based on the NIR CC diagram. Open squares represent the sources with the redder ($H$ - $K$) color (with higher magnitude uncertainties in J-band). Dotted-dashed curve and continuous curve are the ZAMS from \citet{girardi02} corrected for a distance of $\sim$ 2.0 kpc and $A_V$= 0 mag and 2.0 mag, respectively. 
	The diamond, star and triangle symbols shown in NIR CC and CM diagrams represent the massive star BD +45 3216, MSX1 and MSX2, respectively (see Sec. \ref{sec:ion-gas}). The sources with proper motions within 3$\sigma$ 
	of mean values are shown with cross symbols.}  
\label{fig3a}
\end{figure*}

 Mid-IR wavelengths can penetrate dense material and are useful to unravel
the sources in the silhouette of dust layers. In the absence of $Spitzer$ longer wavelength data, we used the WISE data for further characterization of the embedded
 YSO candidates. To remove background
contaminants, specially PAH emitting galaxies and Active Galactic Nuclei (AGN) from the
YSO sample, we used the approach developed by \citet{koenig14}. This selection method uses a series of color and 
magnitude cuts to remove the contaminants such as star-forming galaxies, AGN, 
asymptotic giant branch stars and finally to identify as well as to characterize the young stars. 
To identify the Class {\sc i} and Class {\sc ii} candidate sources, we used 
a sample of remaining point sources  after contamination removal. 
Based on the WISE colors only, the reddest sources are classified as Class {\sc i} if their colors match all of the following criteria:\\
(i) W2 - W3 $>$ 2.0, (ii) W1 - W2 $>$ -0.42 $\times$ (W2 - W3) + 2.2, 
(iii) W1 - W2 $>$ 0.46 $\times$ (W2 - W3) - 0.9, (iv) W2 - W3 $<$ 4.5.

Sources that were not considered as Class {\sc i} candidates, are classified as Class~{\sc ii} candidates if their
colors match all of the following criteria:
(i) W1 - W2 $>$ 0.25, (ii) W1 - W2 $<$ 0.9 $\times$ (W2 - W3) - 0.25, 
(iii) W1 - W2 $>$ -1.5 $\times$ (W2 - W3) + 2.1, (iv) W1 - W2 $>$ 0.46 $\times$ 
(W2 - W3) - 0.9, (v) W2 - W3 $<$ 4.5.

We also identified the YSO candidates using the sources having $WISE$ W1 and W2 magnitudes alongwith the
2MASS $H$ and $K$ magnitudes. These sources are classified as Class~{\sc i} 
candidates if they follow all of the criteria :\\
(i) $H$ - $K$ $>$ 0.0, (ii) $H$ - $K$ $>$ -1.76 $\times$ (W1 - W2) + 0.9, 
(iii) $H$ - $K$ $<$ (0.55/0.16) $\times$ (W1 - W2) - 0.85, 
(iv) W1 $\le$ 13.0, (v) $H$ - $K$ $>$ -1.76 $\times$ (W1 - W2) + 2.55.

The sources obeying all the above conditions except the last one (fifth condition), are classified as Class~{\sc ii} candidates.

The $WISE$ color-color (CC) diagram for the young stars identified using WISE and 2MASS color/magnitude criterion in the Sh2-112 region is shown in Fig. \ref{fig3a} (top-left panel). Open and filled circles represent the Class~{\sc i} and Class~{\sc ii} sources identified based on the WISE / 2MASS $J$, $H$, $K$ magnitudes and colors. In total, we identified 7 Class~{\sc i} and 12~Class {\sc ii} sources in the region.
\subsubsection{NIR color-color space}
YSO candidates emit excess emission in NIR wavelengths which can be assessed based on their 
location in NIR color-color space. 
In Fig. \ref{fig3a} (top-right panel), we show the NIR ($J$ - $H$)/($H$ - $K$) CC diagram constructed using the combined 2MASS and UKIDSS NIR catalog. 
The thin black curve in the CC diagram represents the main-sequence (MS) locus and the thick blue curve is the locus of the 
	giants \citep{bessel88}. The thin green line shows the classical T-Tauri stars (CTTS) locus \citep{meyer97}. 
	The parallel black dashed lines are the reddening vectors drawn from the base of the MS locus, the turning point of the MS locus, 
and the tip of the CTTS locus, respectively. All the magnitudes, colors, and loci of the MS, giants and CTTS are converted
	to the Caltech Institute of Technology system. We have adopted the extinction laws of \citet{cohen81}, i.e., A$_J$ /A$_V$ = 0.265, A$_H$ /A$_V$ = 0.155 and A$_K$ /A$_V$ = 0.090. 
	The sources in the NIR CC diagram are classified into three regions, namely, `F', `T', and `P' \citep[cf.][]{ojha04a,ojha04b}. 
The sources in the `F' region are generally considered as MS/evolved field stars or Class~{\sc iii} YSO candidates (Weak-line T-Tauri Stars), which may be both reddened or unreddened. 
The sources in the `T' region are mainly Class~{\sc ii} YSO candidates \citep[CTTS;][]{lada92} with a large NIR excess and/or reddened 
	early-type MS stars with excess emission in the K-band \citep{mall12}. The sources in the `P' region are Class~{\sc i} YSO candidates 
with circumstellar envelopes. However, we note that there may be an overlap of Herbig Ae/Be stars with the sources in the `T' and ‘P' regions, 
	which generally occupy the place below the CTTS locus in the NIR CC diagram \citep[for more detail; see][]{hern05}. 
	To decrease contamination to our YSO candidates sample, we do not consider sources which were not falling in the `T' or `P' regions within 1$\sigma$ uncertainty in their colors. Adopting this scheme, we have identified a total of 2 Class~{\sc i} and 54 Class~{\sc ii} YSO candidates.
\begin{table*}
\centering
\caption{YSO candidates from 2MASS/UKIDSS and WISE data. The entire table is available in electronic form. }
\tiny
\hspace{-2.0cm}
\begin{tabular}{ccccccccccc}
\hline
Id     & RA & DEC & J$\pm$eJ & H$\pm$eH & $K$$\pm$e$K$& [3.4]$\pm$ & [4.6]$\pm$ &[12]$\pm$ &[22$\pm$&flag\\
	& (J2000) &(J2000)&   &          &         &  e[3.4]    &  e[4.6]    & e[12]    &   e[22] &\\
\hline
	1       &308.43758 & 45.469861 &18.43 $\pm$ 99.99& 17.24 $\pm$ 99.99& 15.12 $\pm$ 0.13 & 11.90 $\pm$ 0.02 & 10.14 $\pm$ 0.02 & 7.76 $\pm$ 0.02 & 4.79 $\pm$ 0.03 &WISE\\
	2       &308.35136 &45.656319 &15.01 $\pm$ 0.04 & 13.56 $\pm$ 0.03& 12.91 $\pm$ 0.03 & 11.96 $\pm$ 0.07 & 11.14 $\pm$ 0.04 & 7.47 $\pm$ 0.06 & 2.45 $\pm$ 0.03   &WISE\\
	3       &308.416255 & 45.618854 &17.58 $\pm$ 99.99 & 15.508 $\pm$ 0.13 & 13.72 $\pm$ 0.05 & 10.53 $\pm$ 0.02 & 9.75 $\pm$ 0.02 & 5.03 $\pm$ 0.01 & 2.04 $\pm$ 0.03&WISE\\
.  &.....&.....&.....&.....&.....&.....&.....&.....&.....\\
\hline
\end{tabular}
\end{table*}
\subsubsection{NIR color-magnitude space}
The NIR CM diagram is a useful tool to identify a population of YSO candidates with IR excess, 
which can be easily distinguishable from the MS stars. 
There may be some YSO candidates that were not detected in the $J$ band. Hence, to identify additional YSO candidates in the Sh2-112 region, we also used the $H$, $K$-bands photometry of the sources 
which do not possess $J$-band photometry or have higher $J$-band magnitude uncertainties ($>$0.1 mag). 
Fig. \ref{fig3a} (Bottom panel) shows the NIR $K$/($H$ - $K$) CM diagram of the 
sources in the region. The dotted-dashed curve and continuous curve 
show the ZAMS loci for a distance of 2.0 kpc with a foreground extinctions of $A_V$ = 0 and 2 mag, 
respectively. NIR excess sources identified based on the NIR CC diagram 
are shown with the open and filled circles. 
As the UKIDSS catalog is deeper than the 2MASS catalog \citep{lucas08}, therefore, most of the faint and redder sources seen in the CM diagram are observed only in the UKIDSS catalog.
In the CM diagram, a low-density gap of sources can be seen at ($H$ - $K$) $\sim$ 0.9 mag, 
therefore, red sources ($H$ - $K$ $>$ 0.9 mag) could be the IR excess sources (red sources and presumably YSO candidates). 
This color criterion is consistent with the control field region where all the 
stars were found to have ($H$ - $K$) $<$ 0.9 mag. A total of 71 sources have been detected as YSO candidates using the $K$/($H$ - $K$) CM diagram. In Fig. \ref{fig3a} (bottom panel), open squares represent the probable IR excess sources identified based 
	on the $K$/($H$ - $K$) CM diagram. Open and filled circles represent the 
	Class {\sc i} and Class~{\sc ii} candidates selected based on the NIR CC 
	diagram. The locations of the YSO candidates in the NIR CM diagram show that a majority of them have $K$ magnitude below 11.5 mag, which corresponds to mass $<$ 3 M$\odot$ \citep[assuming a distance of 2 kpc, an average extinction $A_V$ of $\sim$ 2 mag and using the 1-2 Myr evolutionary tracks 
	of][]{siess2000}.  

	A few YSO candidates selected using the above methods may have an overlap among themselves and hence, the YSO candidates detected 
in the NIR CC and CM diagrams were matched to the WISE YSO candidates. As the resolution of 
the WISE bands is poor compare to the UKIDSS, in a few cases, where there were more than one source within
the matching radius, we considered the closest one as the best match. While matching, if the YSO candidate has different classification in both, priority was given to the classification based on the WISE bands because the longer wavelength magnitude provides more robust information about the Class of the YSO. 8 YSO candidates were found common in the WISE and NIR catalogs of YSO candidates. 
In total, we identified 138 Class {\sc i}/ Class {\sc ii} sources in the whole region. The spatial distribution of the YSO candidates is shown with filled circles in Fig. \ref{khha}. We found 6 H$\alpha$ 
emission stars within HCT FOV and one of them is a Class {\sc ii} source. 
A sample list of the YSO candidates with their magnitudes in NIR and WISE bands
is given in Table 2 and the entire table is available
in electronic form only. 
\subsection{Clustering of YSO candidates}
\begin{figure*}
\centering
\includegraphics[width=0.59\textwidth]{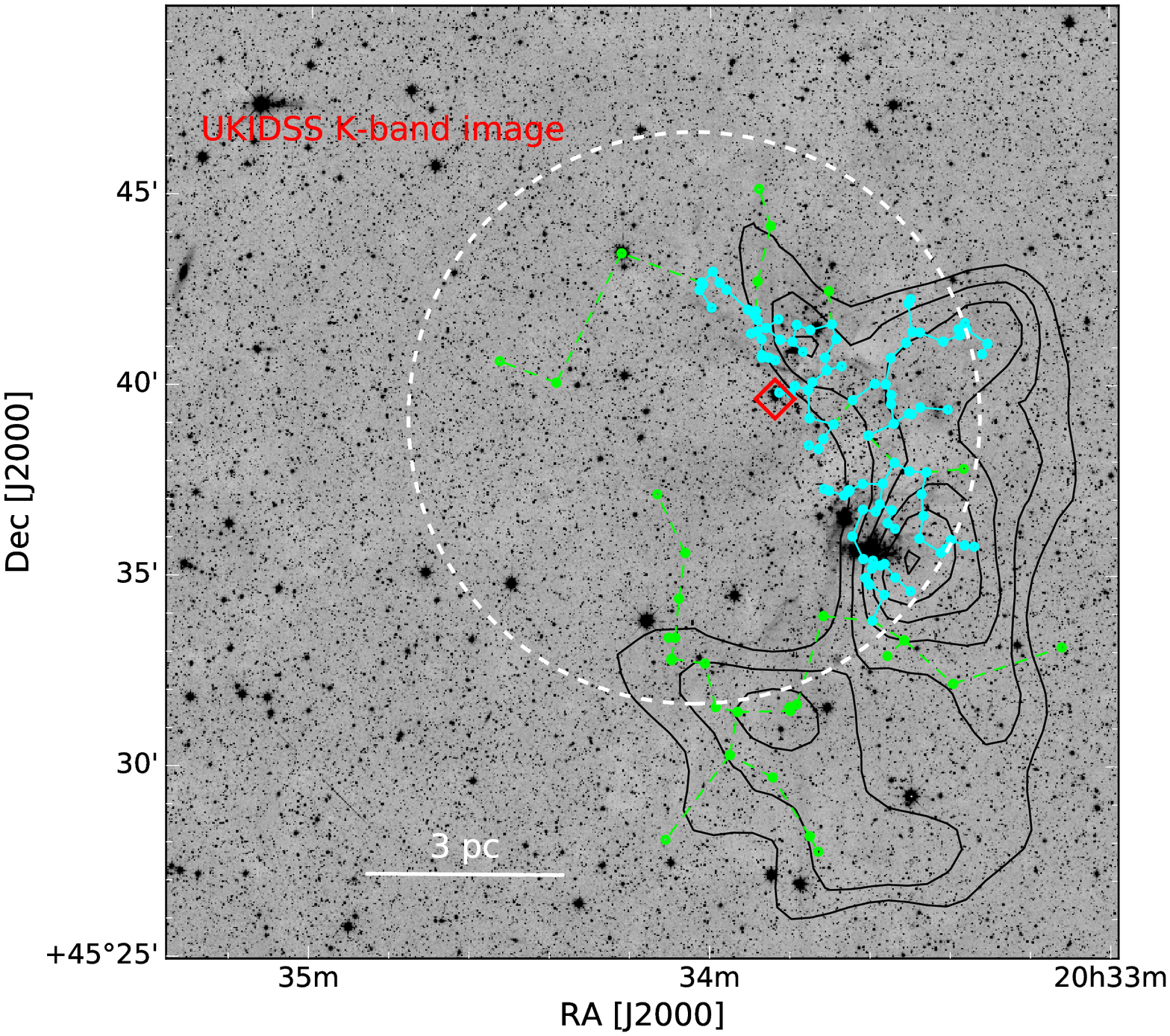}
\includegraphics[width=0.39\textwidth]{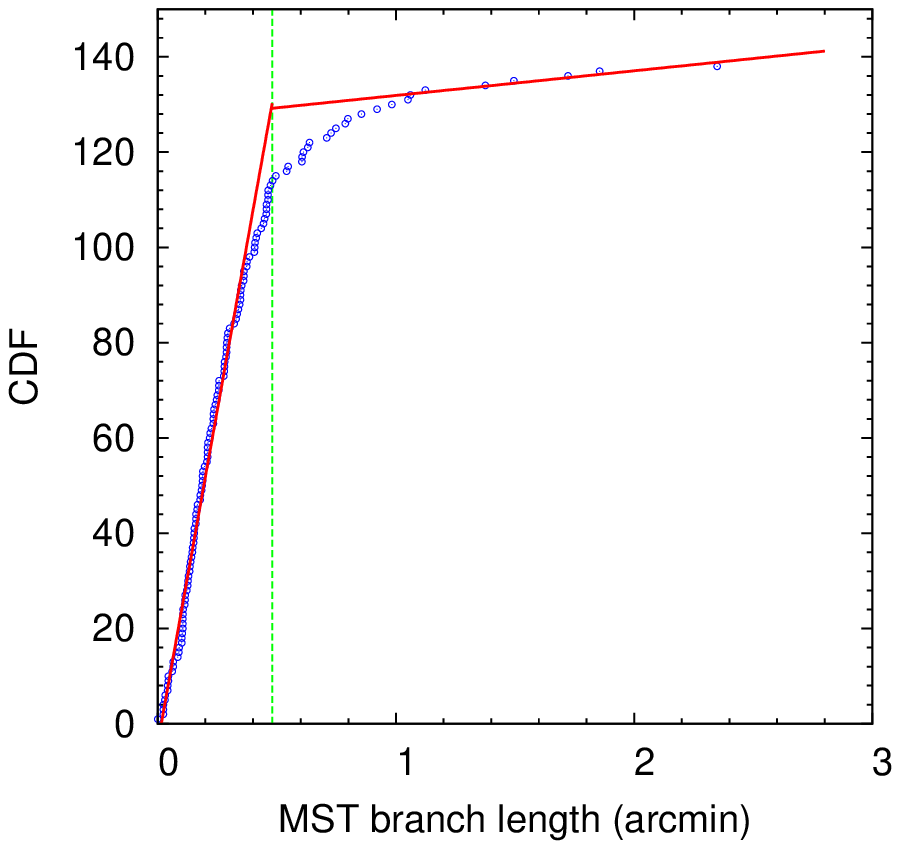}
\caption{Left panel: A UKIDSS $K$-band image of the Sh2-112 region. MST for the identified YSO candidates in the region is also shown with cyan and green lines. The cyan circles 
	connected with cyan lines and green circles connected with dashed green lines are 
	the branches smaller and bigger than the critical length, respectively. 
	The black contours represent the distribution of the dust 
	extinction in the region. 
	The remaining symbols are similar to those in Fig. \ref{fig1}.
Right panel: CDF, used for the critical length analysis of YSO candidates. CDF plot has sorted length values on the X-axis and a rising integer counting index on the Y-axis. The red solid line is a two-line fit for the CDF distribution. The vertical green line is the critical length obtained for the core region.
	}
\label{khha}
\end{figure*}

To study the density distribution of YSO candidates in the Sh2-112 region,
we have  generated the surface density map using the nearest neighbor (NN) 
method \citep[e.g.,][]{guter09},
with a grid size of 15$^{\prime\prime}$ and 6 nearest YSO candidates. 
The YSO surface density map obtained by this method is shown as black contours 
overlaid on the WISE W3, W4 and H$\alpha$ color-composite image of the H{\sc ii} region in Fig. \ref{fig6}b.

The surface density contours clearly reveal 
the sub-clustered and the filamentary distribution of YSO candidates near the western periphery 
presumably due to the fragmentation of filamentary molecular cloud. 
Physical parameters of these sub-clusters, which might have formed in a single star-forming event, 
play a very important role in the study of star formation. 
We further used  minimal spanning tree (MST) based technique to isolate these sub-clusters of YSO candidates from their scattered distribution. 
The MST is defined as the network of lines, or branches, that connect a set of points together such that
the total length of the branches is minimized and there are no closed loops.
This technique efficiently isolates
the sub-structures without any type of smoothening and bias regarding the shapes of the distribution 
\citep{guter09}. This method  also preserves the underlying geometry of the distribution 
\citep[e.g.,][]{cartwright04,schmeja06,guter09,sharma16}. 

To obtain the MSTs for the YSO candidates in the region, we used the approach as suggested by \citet{guter09} 
and \citet{sharma16}. 
In Fig. \ref{khha} (left panel), we have overplotted the derived MSTs for the location of YSO candidates in the Sh2-112. 
The circles and lines denote the positions of the YSO candidates and the MST branches, respectively. 
The MST distribution also clearly shows the sub-clusters near the western periphery of the H{\sc ii} region. 
To isolate these sub-clusters of YSO candidates, we determined the critical 
branch length of the MST by plotting their cumulative distribution. 
The resultant cumulative distribution function (CDF) in 
Fig. \ref{khha} (right panel), clearly shows a three-segment curve; a steep-sloped segment at short spacings, a transition segment that approximates the curved character of the intermediate length spacings, and a shallow part. 
We have fitted two lines 
in the shallow and steep segments and extend these two to connect together. We adopted the intersection point between these 
two lines as the MST critical branch length \citep[see also,][]{guter09,chav14,sharma16}. 
The YSO sub-clusters were then isolated from the lower density distribution by 
clipping MST branches longer than the critical length found above. 
The value of the critical branch length for the sub-clusters is 0.48 arcmin which is shown as a vertical line. 
Cyan circles and MST connections in Fig. \ref{khha} (left panel) represent the locations of YSO candidates in the 
sub-clusters identified using the above procedure.

\subsection{Optical Morphology of the region}
\begin{figure*}
\centering
\includegraphics[width=0.535\textwidth]{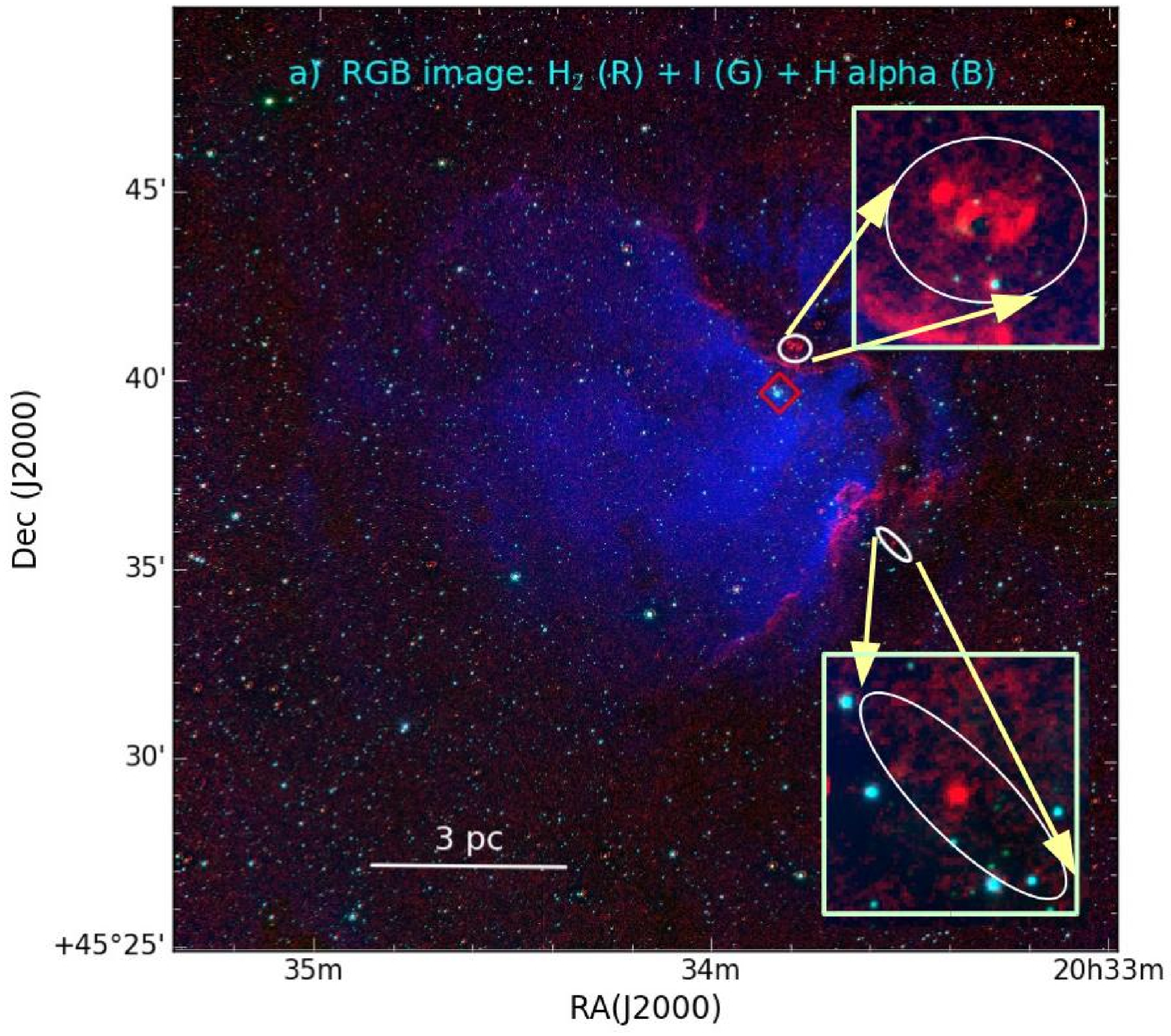}
\includegraphics[width=0.457\textwidth]{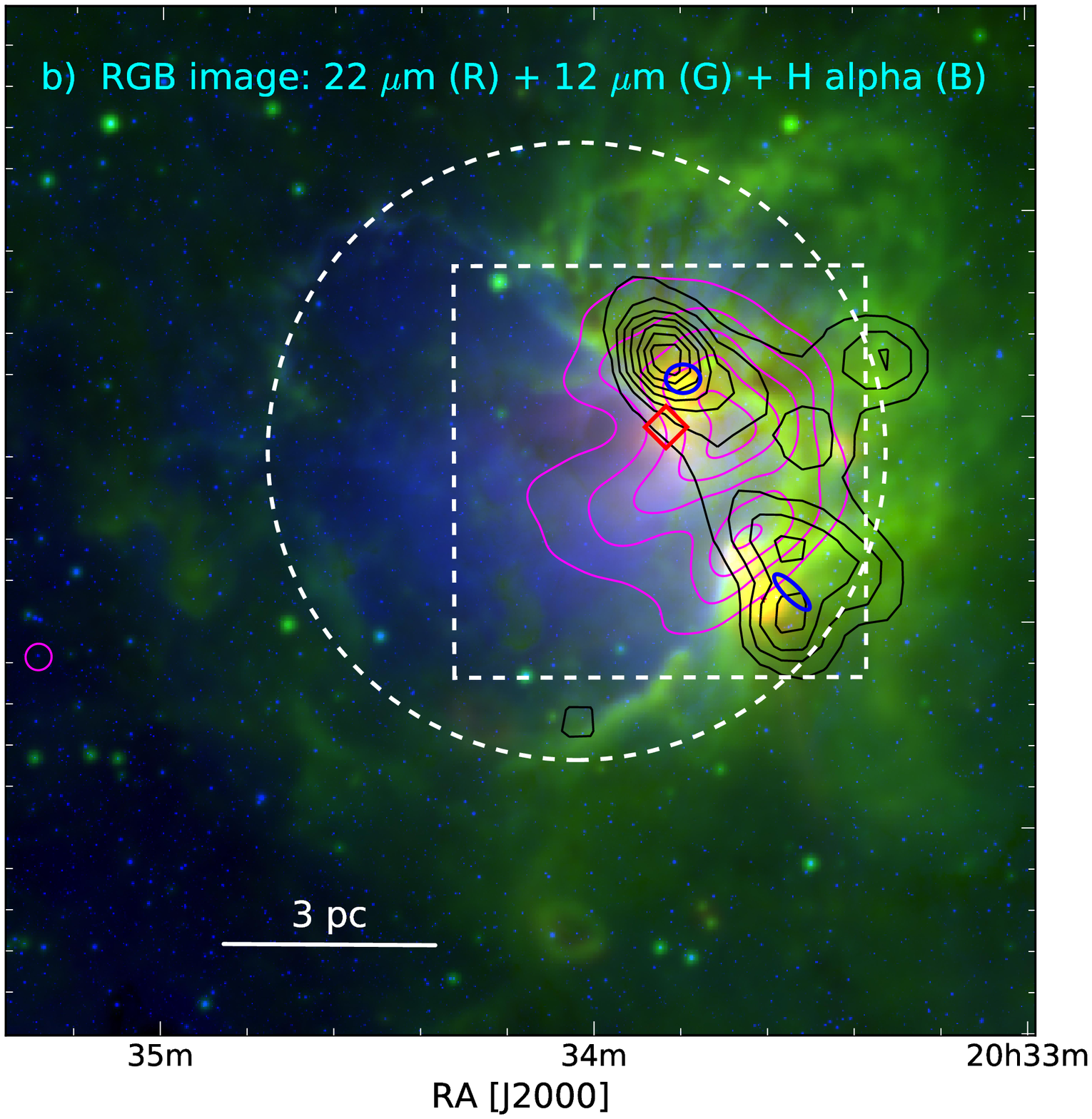}
	\caption{(a) A color composite image of the Sh2-112 region generated from 
	the H$_2$ (red), I-band (green) and H$\alpha$ (blue) images. 
	 Ellipses show the locations of the H$_2$ knots. 
	 (b) A color composite (WISE 22 $\micron$: red, 12 $\micron$: green and KPNO H$\alpha$ : blue) image of the Sh2-112 region. The magenta contours show the distribution of the GMRT 610 MHz radio continuum emission starting at a level of 50 mJy/beam to 
	the peak value of 330 mJy/beam with an increment of 50 mJy/beam. The 
	black contours represent the surface density distribution of the YSO candidates in the region. 
	Other symbols are similar to those in Fig. \ref{fig1}.}
\label{fig6}
\end{figure*}

The H$\alpha$ emission from the region traces the possible extension of the H{\sc ii} region which can be seen in blue color in Fig. \ref{fig6}a. 
Though the H{\sc ii} region, in general, represents the spherical morphology, we notice that the massive star 
is not at the center of the H{\sc ii} region and H$\alpha$ emission is extended only towards
 the eastern side of the massive star. There are dark features visible against the bright H$\alpha$ emission of the H{\sc ii} region. An elongated obscuring
dark lane extending approximately in the north-west to south-west direction
near the western periphery is seen. This obscuring dark lane suggests the probable presence of the molecular cloud. 
The south-west part is brighter and has a sharp arc-like structure in the west direction,
giving an impression of a possible ionization front (IF) whereas
the northern part has an extended and very filamentary structure with 
a stunning view of the dust content, indicating that a significant amount of dust is concentrated
in the filaments across the region. This structured feature
suggests an inhomogeneous spatial distribution of the interstellar
matter.
 In addition, we notice patchy faint H$\alpha$ nebulosity towards the west of the H{\sc ii} region. 
\begin{figure*}
\centering
\hspace{-0.0cm}
\includegraphics[width=0.5\textwidth]{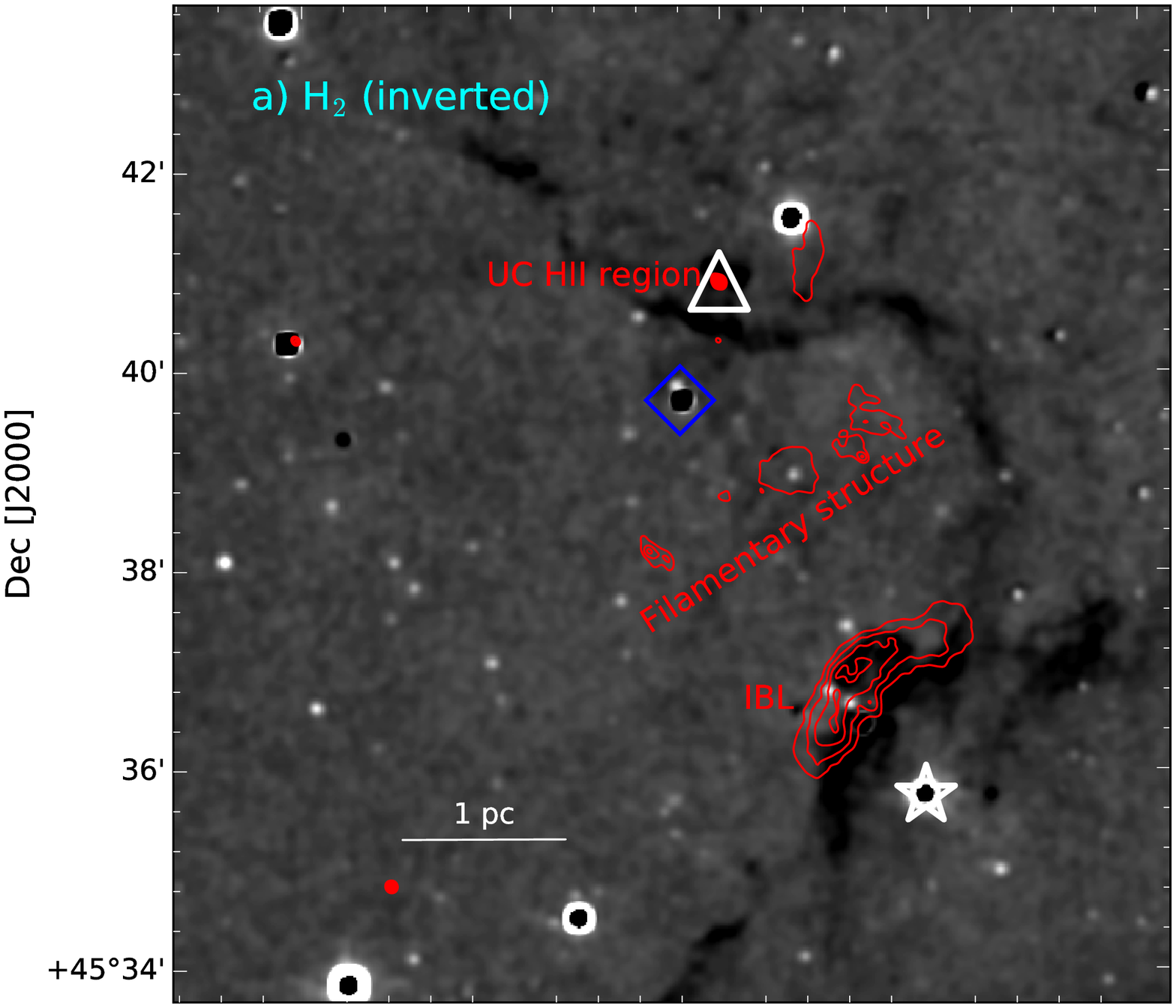}
\hspace{0.35 cm}
\includegraphics[width=0.46\textwidth]{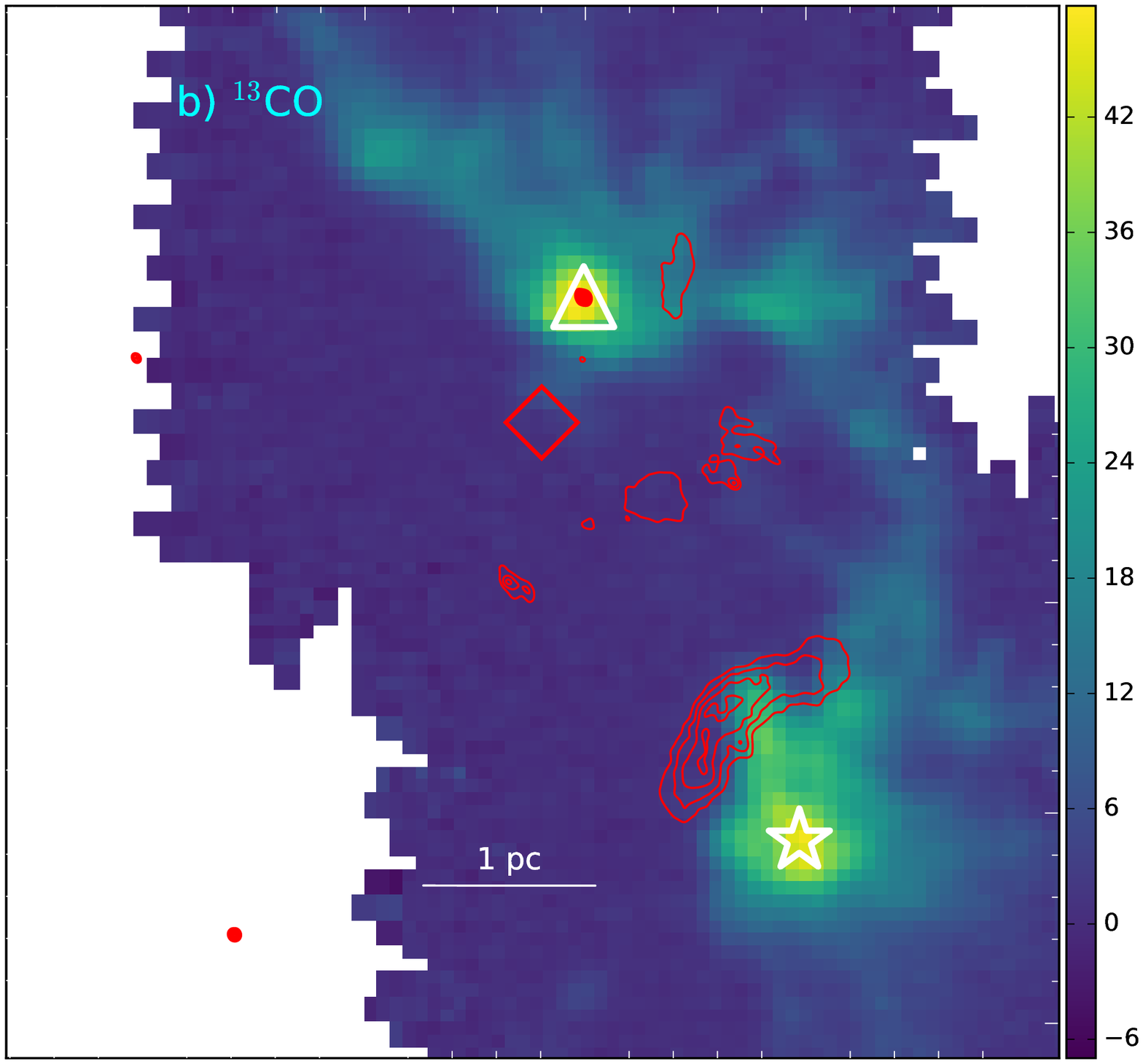}
\includegraphics[width=0.53\textwidth]{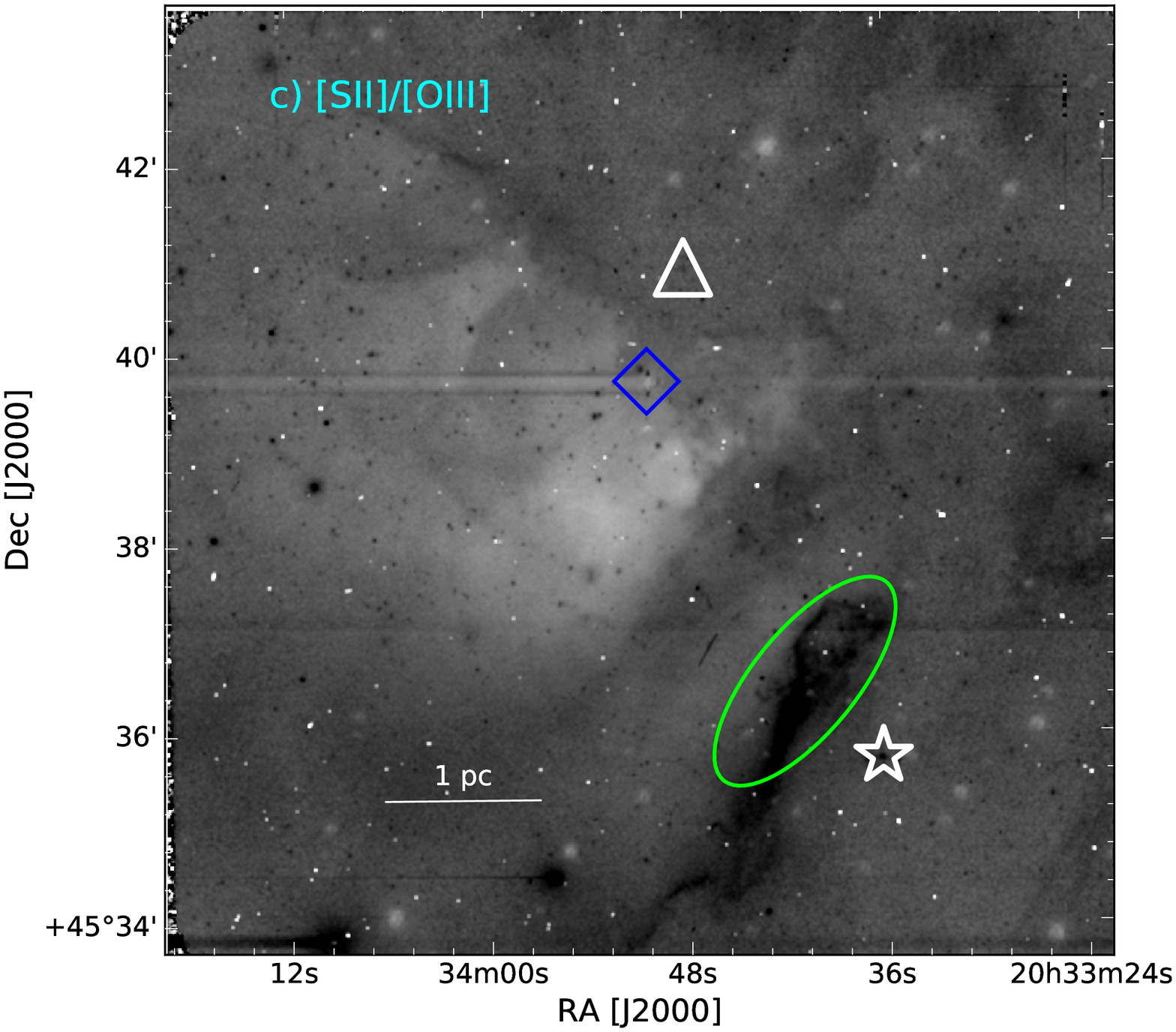}
\includegraphics[width=0.455\textwidth]{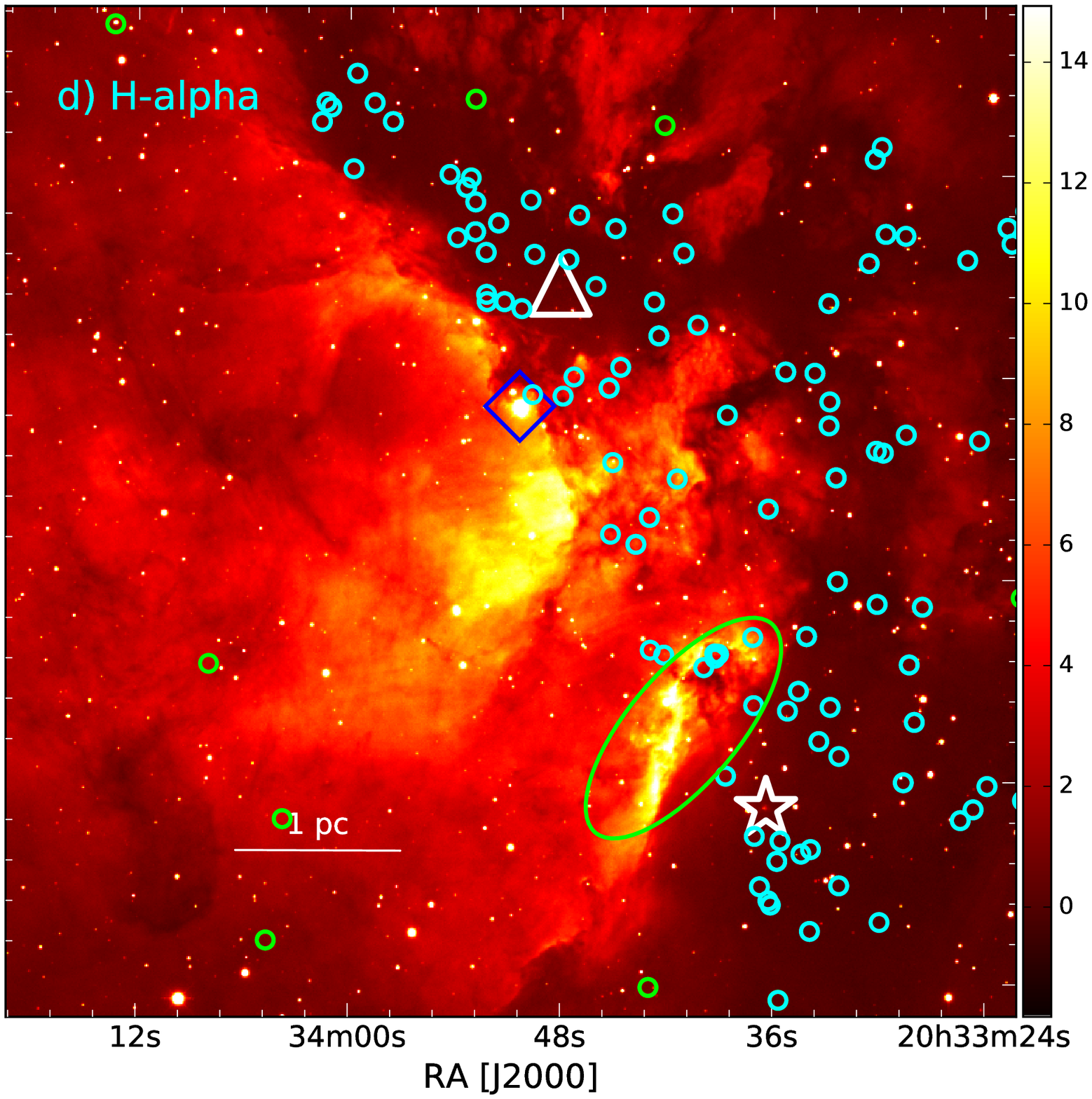}
	\caption{A zoomed-in view of the square region ($\sim$ 10$^\prime$ $\times$ 10$^\prime$) shown in Fig. \ref{fig6}b.
	Panel (a): H$_2$ inverted grayscale map, Panel (b): $^{13}$CO intensity map, Panel (c): [S{\sc ii}]/[O{\sc iii}] ratio greyscale map,
	Panel (d): H$\alpha$ intensity map, of the Sh2-112 region. The distribution of 1280 MHz emission obtained from the 
	GMRT observations is shown as red contours at the level starting from 0.3 mJy/beam to the peak value of 23 mJy/beam with an increment of 0.125 mJy/beam. 
	The circles in panel (d) represent the spatial distribution of the young stars. 
	Star and triangle symbols represent MSX sources MSX1 and MSX2, respectively.
	Ellipse in the lower panel shows the location of the IBL. The remaining symbols are similar to those in Fig. \ref{fig1}.}
	\label{ext}
\end{figure*}
\subsection{Distribution of shocked H$_2$ emission}
The high-excitation molecular gas can be probed using the NIR $H_2$ line
emission at 2.12 $\micron$. Continuum-subtracted H$_2$ line emission image was obtained by removing the 
 off-line emission using the K-band image. For this, first, the image with the 
 best seeing was degraded  by convolution
with a Gaussian function  using the $IRAF$ task $psfmatch$, to match the width 
of the PSF in different images. For the K-band image, a flux scaling factor was
derived by comparing the counts of several stars in both images, 
and then an H$_2$ emission image is obtained by subtracting the K-band scaled image from the H$_2$ image. 

In Fig. \ref{fig6}a, the distribution of a continuum-subtracted H$_2$ emission in the Sh2-112 is shown with red color. In the continuum-subtracted H$_2$ image, compact features with a combination of positive
and negative valued features are the residuals of continuum subtraction of point sources. In
addition to those features, values lower than the background are observed. 
This is due to the continuum whose flux ratio between the H$_2$ narrow
band and K-band filters is different from that of foreground stars. 
Figure \ref{fig6}a reveals diffuse H$_2$ emission along the western border
of the H{\sc ii} region, as well as notable features at various
locations of the nebula. The diffuse H$_2$ emission can arise either 
due to the UV fluorescence from the massive star and thus can trace 
PDR or can be due to collisional excitation by the shocks from outflows originating from nearby YSO candidates \citep[e.g.,][]{chrys92}. The PDR can also be traced by the PAH emission in the WISE 12 $\micron$ band. However, the close resemblance of the H$_2$ emission with the 12 $\micron$ emission (see Fig. \ref{fig6}),
suggests that H$_2$ emission features at the western periphery of Sh2-112 are
more likely caused by the excitation from UV photons. Also, the H$_2$ emission shows rim or
arc-like morphology, with either the rim facing towards
the ionizing source or the curvature of arc appears to be created by the UV photons from 
the ionizing star. 
However, high-resolution and high sensitivity molecular line observations
are needed to shed more light on this. 
The morphology of H$_2$ is different compared to the ionized gas emission traced by H$\alpha$ image. H$_2$ is seen at the edges of the ionized gas emission while there is little or no emission in the central region (see Fig. \ref{fig6}a). This indicates the presence of highly excited molecular gas around the western periphery of the H{\sc ii} region. In addition to the filamentary structures, a few H$_2$ knots are also visible in this region (see the features surrounded by ellipses in Fig. \ref{fig6}). 

The massive star located in this region and its high energy feedback might be responsible for 
H$_2$ and PAH emissions and the  observed arc-like morphology of gas and dust.
\subsection{Distribution of the ionized gas}
	\label{sec:ion-gas}
As the ionized gas of the H{\sc ii} region emits in the radio continuum
due to free-free (Bremsstrahlung) radiation, the radio continuum images can be used to trace the 
distribution of the ionized gas in the 
H{\sc ii} region. In Fig. \ref{fig1}, the distribution of 21 cm radio continuum emission from CGPS is shown in blue. 
The distribution of continuum emission suggests a complex morphology 
of the ionized gas emission. The location of a previously 
characterized O8V star (BD +45 3216) does not appear near the peak of radio continuum emission. The bulk of the ionized emission is located towards the western periphery. 
To better understand this region, we have further used the GMRT data at 610 MHz and 1280 MHz, 
which has better resolution and sensitivity than the CGPS 1420 MHz data.

The GMRT radio contours
at 610 MHz (magenta contours) are overlaid on a color-composite image (WISE 22 $\micron$: red; 12 $\micron$: green; H$\alpha$: blue) 
of the Sh2-112 H{\sc ii} region in Fig. \ref{fig6}b. 
Furthermore, the correlation of the warm dust and ionized emission is also evident, 
which has generally been found in the H{\sc ii}
regions \cite[e.g.,][]{deharveng10,paladini12}. In addition to the extended emission, the embedded 
stellar contents are seen near and around the arc-like structure. 

Fig. \ref{ext} shows the zoomed-in view of the square region ($\sim$ 10$\arcmin$ $\times$ 10$\arcmin$) shown in Fig. \ref{fig6}b. 
In Fig. \ref{ext} (top-left panel), we overlaid the GMRT 1280 MHz contours on the inverted H$_2$ image. It 
is clear from Fig. \ref{fig6}b that the ionized emission associated with this H{\sc ii} region 
displays a complex morphology at 610 MHz with a steep intensity 
gradient towards the west. A faint, broad and diffuse emission is seen towards the north-west. 
The 610 MHz emission contours at the western periphery seem to be composed of 
three components in the 1280 MHz emission; a clearly visible arc-shaped structure 
towards the south-west boundary, an extended filamentary feature above the arc-shaped 
structure and a compact source towards the north of the massive star BD +45 3216. 
This compact radio source at 1280 MHz could possibly be a compact/ ultra-compact (UC) H{\sc ii} region.

In this UC H{\sc ii} region, the free-free emission at 1280 MHz is assumed to be optically thin and considering that the
region is in photo-ionization equilibrium, the
total number of ionizing photons (N$_{UV}$) from the massive source is obtained by 
using the integrated radio flux within the region (S$_\nu$), following the equation given in \citet{morgan04} i.e., 
\begin{equation}
N_{UV} = 7.7 \times 10^{43}S_\nu D^2 \nu^{0.1}
	\end{equation}
where S$_\nu$ is in mJy, D is the distance to the source in kpc and $\nu$ 
is the frequency of the observation in GHz. 
The estimated Lyman continuum flux for the UC H{\sc ii} region is 
10$^{47.1}$ photons s$^{-1}$ for a distance of $\sim$ 2 kpc, which 
corresponds to a single ionizing star of spectral type 
of B0 - B0.5 V (see Table 2 of \citet{panagia73}). 
As the optical images do not show any source corresponding to the radio continuum peak, however, 
longer wavelength images reveal presence of a source inside it, suggesting that the massive star is still embedded within the molecular material.

An extended filamentary feature above the arc-shaped structure (see Fig. \ref{ext}, top-left panel) also appears as the enhanced H$\alpha$ emission 
as seen in Fig. \ref{ext} (bottom-right panel). However, this feature is not 
visible in the WISE 12 $\micron$ image and seems to be due to the high intensity of 
ionized gas emission.

The location of a noticeable arc-like feature in radio emission contours matches very well with the curved bright emission feature in H$\alpha$ image. This 
positional and morphological correlation of arc-like structure in radio and optical images suggests that the radio emission may be originating from the 
ionized boundary layer (IBL) located at the border of the dark molecular cloud. IBL is a recombination layer that 
develops on the side of the cloud facing the ionizing star when the surface 
of the molecular cloud becomes ionized. 
Thus, the radio emission map at 1280 MHz is utilized  to estimate 
the ionizing flux ($\phi$$_p$) impinging on the curved rim-like feature and the electron density (n$_e$) within the IBL. 

We calculated the ionizing flux and the electron density using the equations given in \citet{lefloch97} and \citet{thompson04} :  
\begin{equation}
	{\phi}_p = 1.24 \times 10^{10} S_{\nu} T_{e}^{0.35} {\nu}^{0.1} {\theta}^{-2}
\end{equation}
\begin{equation}
	n_{e} = 122.41 \times \sqrt{\frac{S_{\nu} T_{e}^{0.35} {\nu}^{0.1} {\theta}^{-2}}{\eta R}},
\end{equation}
where T$_e$ is the effective electron temperature of the ionized gas in K, $\nu$ is the frequency
of the free-free emission in GHz, $\theta$ is the angular diameter
over which the emission is integrated in arcsec. 
R and $\eta$ are the radius of the cloud in pc and effective thickness of the 
IBL as a fraction of the cloud radius, respectively. 
$\eta$ was found to be vary in the range 0.1 - 0.2 of the cloud radius 
and is primarly dependent upon the ionising flux and the cloud curvature \citep{bert89}.  Here, we have taken $\eta$ = 0.2, which implies that the derived electron density is strictly a lower limit. The value of R is taken as $\sim$ 0.8 pc. We derived the photon flux value ($\phi$$_p$) of 2.34 $\times$ 10$^{10}$ /cm$^2$/s which 
is larger than the ionizing flux predicted (1.25 $\times$ 10$^{10}$ /cm$^2$/s) at this location due to an O8 V star. This excess of the measured ionizing flux may be due to another embedded radio source. 
The WISE 22 $\micron$ image reveals extended emission from a point source near the 
 IBL, 
 further strengthening the possibility of an embedded radio source near the IBL. 
 We obtained an electron density of $\sim$ 413 cm$^{-3}$ which is greater than the critical value of $\sim$ 25 cm$^{-3}$ above which an IBL is able to develop around a molecular cloud \citep[see][]{lefloch95}. 

We looked for the Midcourse Space Experiment (MSX) point sources towards the 
UC H{\sc ii} region and IBL. We found a source MSX 6C 083.7071+03.2817 (hereafter `MSX1') $\sim$ 6 $^\prime$ away from the 
IBL and another source MSX 6C 083.7962+03.3058 (hereafter `MSX2') near to the UC H{\sc ii} region. 
The locations of the MSX1 and MSX2 are shown in Fig. \ref{ext} as star and triangle symbols, respectively. 
We also looked for the 2MASS counterparts of the MSX1 and MSX2. The locations of these two sources are 
shown in the NIR CC and CM diagrams (see Fig. \ref{fig3a}). 
We note that the magnitude uncertainty associated with the 2MASS magnitudes of MSX2 is higher. The 
$H$/($H$ - $K$) CM diagram shows that these sources (MSX1 and MSX2) could have foreground extinction, A$_V$, of $\sim$ 21 mag and 19 mag, respectively. The NIR CM diagram shows that the MSX2 may be of B1-B2 spectral type which is comparable to that obtained based on the radio continuum emission at 1280 MHz.

\citet{urqu09} also observed a compact VLA source at 6 cm radio continuum emission toward MSX2 and designated it a compact H{\sc ii} region. \citet{cooper13} carried out NIR  spectroscopy of massive YSO candidates in several SFRs including MSX1. They classified 
MSX1 as a massive YSO candidate with a luminosity of $\sim$ 5.7 $\times$ 10$^3$ L$\odot$ embedded in extinction of  
A$_V$ $\sim$ 26-36 mag. Towards the further in the direction of the western periphery of the Sh2-112 H{\sc ii} region
($\sim$ 4 arcmin away from the MSX1), another compact radio continuum source can be seen, which was also observed 
by \citet{urqu09} at 6~cm VLA observations. 

The dynamical age of the H{\sc ii} region can be estimated using the equation given in \citet{dyson80}:
\begin{equation}
t_{dyn} = \left(\frac{4\,R_{s}}{7\,c_{s}}\right) \,\left[\left(\frac{R_{HII}}{R_{s}}\right)^{7/4}- 1\right]\\
\end{equation}
where c$_{s}$ is the isothermal sound velocity in the ionized gas \citep[c$_{s}$ = 10 km s$^{-1}$;][]{bisbas09},
R$_{HII}$ is the radius of the H{\sc ii} region, and R$_{s}$ is the radius of the Str\"{o}mgren sphere and is calculated using the equation :\\
\begin{equation}
	R_{s} = (3 N_{UV}/4{\pi} n^2 {\alpha}_{B})^{1/3}
\end{equation}
where the radiative recombination coefficient $\alpha_{B}$ =  2.6 $\times$ 10$^{-13}$ (10$^{4}$ K/T)$^{0.7}$ cm$^{3}$ s$^{-1}$ \citep{kwan97}. N$_{UV}$ is the number of UV photons/s, and `$n$' is the initial particle number density of the ambient neutral gas. 
Considering a typical value of N$_{UV}$ for an O8V star from \citet{panagia73}, and R$_{HII}$ as 7$\arcmin$.5, the calculated dynamical age might vary substantially depending on the initial density.  
The dynamical age of the H{\sc ii} region Sh2-112 varies from 1.6 Myr to $\sim$ 5 Myr for a range of ambient density ($n$) from 10$^{3}$ cm$^{-3}$ 
to  10$^{4}$ cm$^{-3}$. 
We also calculated the dynamical age of the UC H{\sc ii} region. Assuming $n$ (=10$^{5}$ cm$^{-3}$), we estimated the dynamical age of the UC H{\sc ii} region as 0.01 Myr.

\subsection{Distribution of Molecular Gas and Cold Dust}

The molecular hydrogen gas in the SFRs can be detected via CO observations. 
\citet{dobashi94} undertook CO observations of the whole Cygnus complex 
including H{\sc ii} region Sh2-112. Their low-resolution CO maps show the presence of the CO emission in the direction of Sh2-112. In absence of high-resolution CO observations, 
dust content via  extinction measurement is a direct and reliable 
tracer of the hydrogen content of the molecular clouds. 
The  extinction measurements can be done by measuring the color-excess in the IR wavelengths \citep{lada94}.
To generate the dust extinction map for the Sh2-112 region, we used our NIR catalog  and 
followed the method discussed in \citet{panwar14}. The region is divided into
a number of small cells and  
looked for the ($H$ - $K$) colors of 20 nearest stars to the cell center. 
We computed the color excess E = ($H$ - $K$)$_{obs}$ - ($H$ - $K$)$_{int}$,
where ($H$ - $K$)$_{obs}$ is the observed median color in a cell, and
($H$ - $K$)$_{int}$ is the intrinsic median color estimated from the colors
of supposedly unreddened stars. 
We calculated the extinction in K-band (A$_K$), within each cell using the relation A$_K$ = 1.82 E \citep{flaherty07}. Further, the A$_V$ value is 
obtained using the relation A$_V$ = 15.87 E. 
The extinction map for the H{\sc ii} region is plotted in Fig. \ref{khha} (left panel) as thick black contours. The contour starts with 25\% of the peak value with an increment of 15\%. The average spatial resolution 
of the extinction map is $\sim$ 30 arcsec (which converts to 0.3 pc at 2 kpc distance). However, the resolution of the extinction map depends on the surface 
density of the stars and is higher for the regions with high surface density and lower for the regions with 
low surface density. The main factors contributing to the uncertainty in the extinction measurement are the random error in the measurement of ($H$ - $K$) colors and the systematic error due to the adopted extinction law. 
The extinction map clearly indicates that the dust is present towards the western periphery of the H{\sc ii} region. The deficit of the dust content and subsequently  
molecular material towards the eastern part may be due to escaping ionized
hydrogen gas in this direction.

As the distribution of molecular gas in SFRs is clumpy and in general, the $^{13}$CO line is more optically thin compared to the $^{12}$CO line. Therefore, the $^{13}$CO line data can trace dense condensation and its associated velocity better than $^{12}$CO. JCMT $^{13}$CO image of the western periphery of the H{\sc ii} region (near the source MSX1 and MSX2) integrated in 
the [-16, -1] km s$^{-1}$ velocity range is shown in Fig. \ref{ext} (top-right panel). 
This region also appears to be associated with  H$_2$ and/ or PAH features (see Fig. \ref{fig6}), suggesting a close interaction between ionized and molecular gas in the complex. 
As NH$_3$ is generally used as a high-density gas tracer, \citet{urqu11} observed the MSX 
sources MSX1 and MSX2 for NH$_3$ (1,1), (2,2), (3,3) and water maser emissions 
and detected NH$_3$ emission 
towards both MSX sources whereas water maser was observed only towards the MSX1. 
\citet{maud15b} have also detected massive molecular outflows towards MSX1 and found that the dynamical time scale of redshifted and blueshifted lobes of the outflows is of the order of 4 $\times$ 10$^4$ years. 
They also suggested that the MSX2 is also probably associated with the molecular outflows. 
$^{13}$CO observations by \citet{urqu09} revealed a component of V$_{LSR}$ $\sim$ -3.2 km/s associated with MSX1 whereas two components of 
V$_{LSR}$ -11.6 km/s and -3.7 km/s are associated with MSX2. \citet{maud15a} also 
observed strong C$^{18}$O emission towards these sources. Thus, western side 
of the H{\sc ii} region seems to possess molecular gas, dense cores and associated with recent/ongoing star formation activity.

\begin{figure*}
\centering
\includegraphics[width=0.45\textwidth]{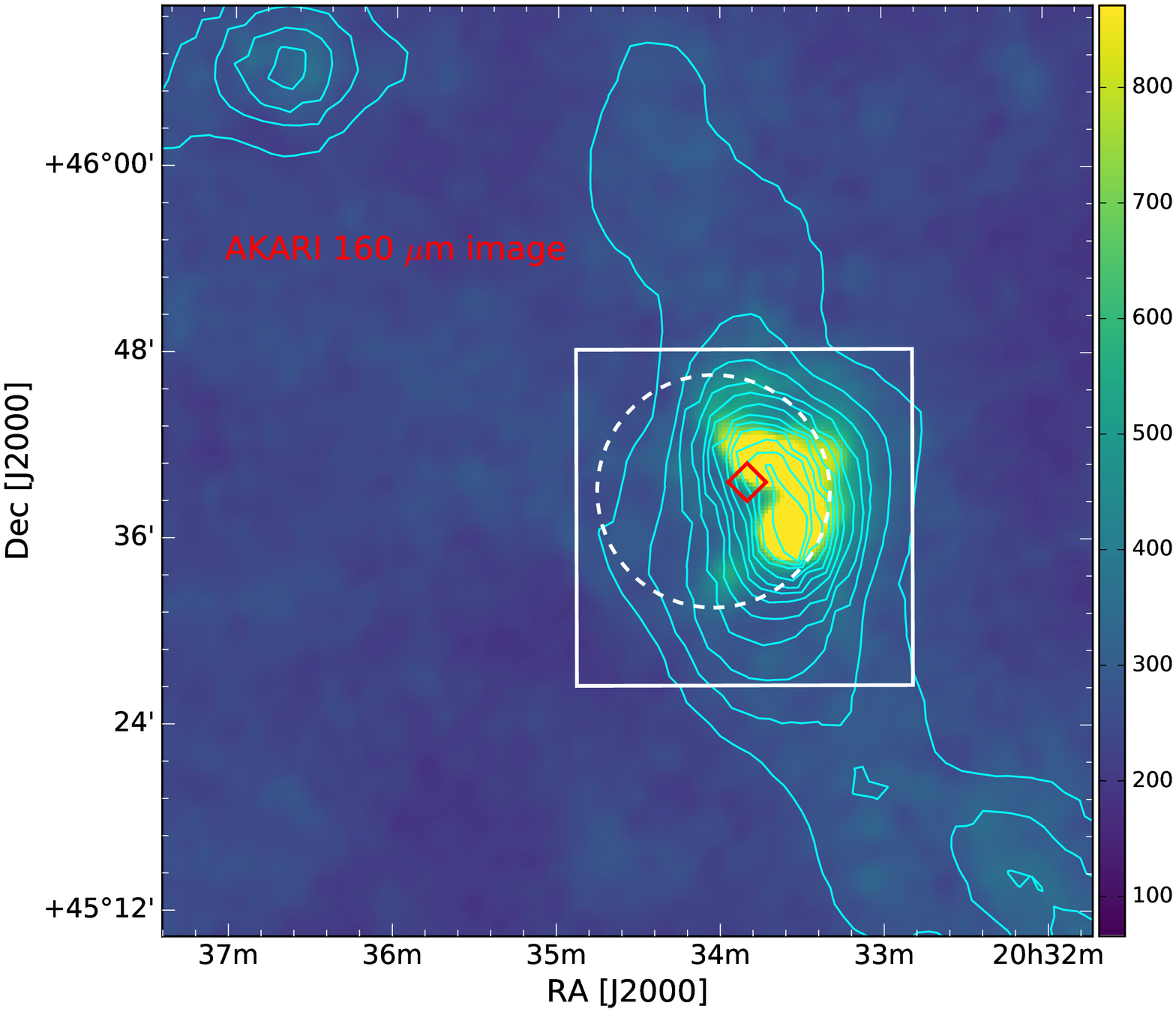}
\includegraphics[width=3.5in, height=2.8in, trim = {0.0in 2.25in 0.25in 2.25in}, clip]{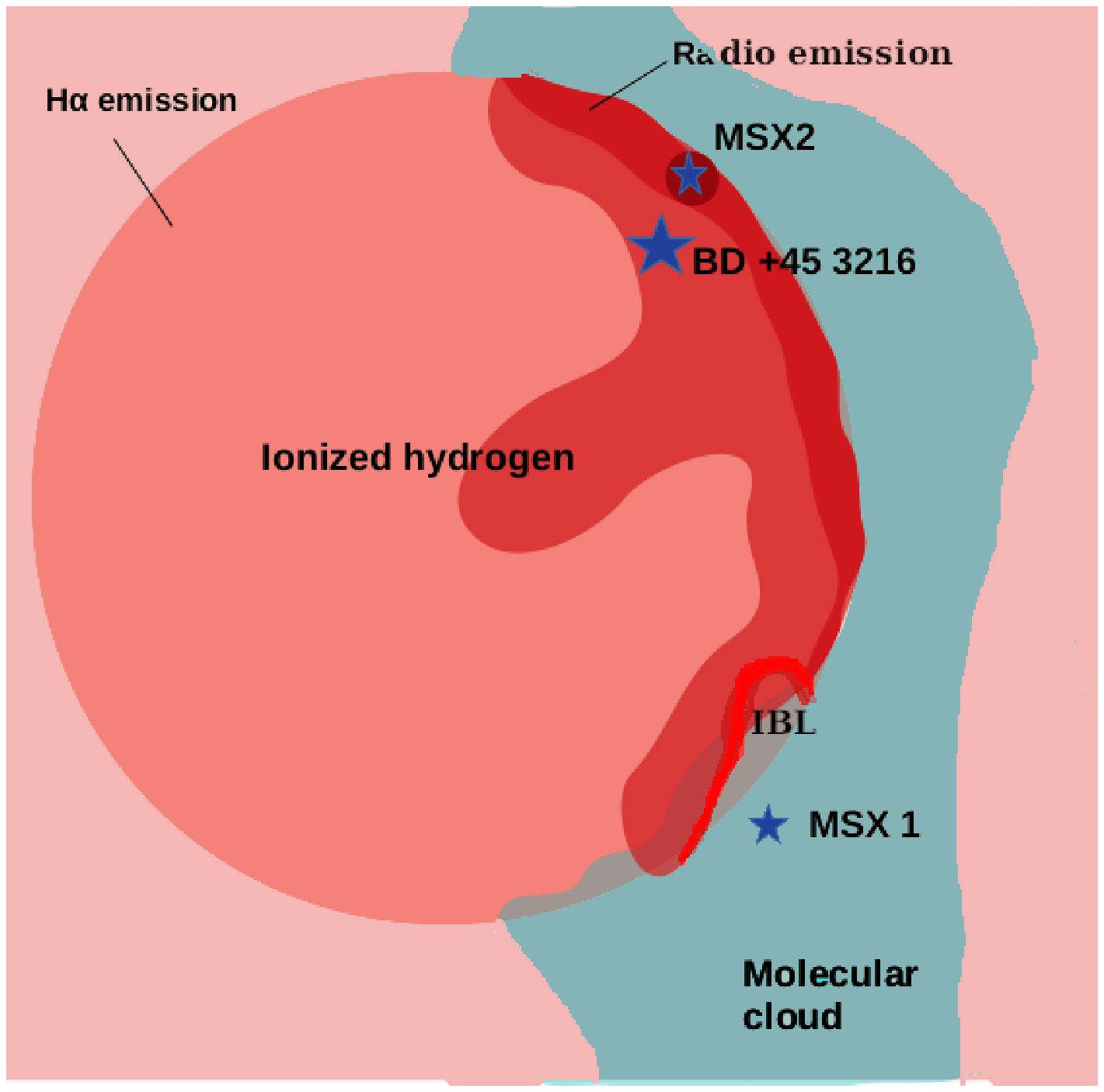}
	\caption{Left: AKARI 160 $\mu$m image superimposed by Planck 353 GHz contours. The contour starts from 
	0.033 mJy/Sr with a step size of 0.005 mJy/Sr.
	Right: Cartoon diagram representing the morphology of the H{\sc ii} region Sh2-112 within the white box shown in the left panel. }
	\label{morph}
\end{figure*}

\section{ Discussion}
\subsection{Sh2-112: a candidate blister {\rm H}{\sc ii} region}
An H{\sc ii} region develops a spherical morphology if the massive star forms in a uniform molecular cloud. 
However, if the massive star forms close to the edge of a molecular cloud, the expanding IF/SF reaches the edge of the cloud quickly. Consequently, the H{\sc ii} region opens in the direction away from the cloud edge \citep{tenorio79,henney05,krumholz09}
as ionized gas would be less confined and could
stream out into the lower density ISM easily. It rapidly propagates through
the low-density ISM, followed by a strong shock
driven by the higher pressure ionized cloud material. At the same
time, a rarefaction wave propagates back into the ionized cloud,
giving rise to a ``blister-type'' H{\sc ii} region showing the champagne flow of ionized material \citep[see, e.g.,][]{duronea12,2007MNRAS.379.1237M}. 
The Sh2-112 H{\sc ii} region represents a spherical shell of a diameter $\sim$ 15 arcmin ($\sim$ 8.7 pc,  
at the distance of $\sim$ 2 kpc).  $^{13}$CO and extinction maps suggest the presence of molecular gas 
near the western periphery of the H{\sc ii} region. This molecular material 
also appears as a highly extinct dark lane in the optical 
H$\alpha$ image (cf. Secs. 3.7 and 3.8), where most of the YSO candidates identified in the present work 
are located.
As described in Secs. 3.6 \& 3.7, the H$_2$ and the PAH emission, indicating the 
location of the PDR, encircles the western border of the H{\sc ii} region. 
A curved morphology of the radio continuum emission in the 
western periphery further suggests a stalling effect for the ionized gas
against the PDR in that direction. 

As shown by \citet{pelle12}, the [S {\sc ii}]/[O {\sc iii}] ratio map can be used to 
distinguish density-bounded, radiation bounded and blister-type H{\sc ii} 
regions. \citet{pelle12} found that the optically thick nebula with highly 
ionized gas (low [S {\sc ii}]/[O {\sc iii}] ratio; appears dark in ratio map) is usually surrounded by an ionization transition zone with 
higher [S~{\sc ii}]/[O~{\sc iii}] ratio (appears lighter grayscale). 
We made a [S~{\sc ii}]/[O~{\sc iii}] ratio map of the central part ($\sim$ 10$^\prime$ $\times$ 10$^\prime$) of the Sh2-112 H{\sc ii} region, which is shown in Fig. \ref{ext} (bottom-left panel). The ratio map consists of bright and dark features. We notice a pronounced ionization transition zone (bright region) towards the south of the 
massive star, where H$\alpha$ emission is traced.
This ionization transition zone appears to be surrounded by a relatively diffuse feature in light gray.
In the south-west direction, the ratio of [S {\sc ii}]/[O {\sc iii}] reveals the presence of ionized gas (dark feature) in the region, which is spatially correlated with the IBL. 
The distribution of the ionized hydrogen, H$_2$ and $^{13}$CO emission indicates that the IBL structure is seen edge-on. An example of this kind of edge-on geometry is 
the Orion bar which is located in a blister-type H{\sc ii} region, the Orion Nebula \citep{goico18}. 
In our case, the bright S {\sc ii} emission could be actually on the surface of the H{\sc ii} cavity. 
Therefore, from the above analysis, it seems that the Sh2-112 is an example of a blistered H{\sc ii} region which is 
ionization bounded on the high-density sides (west direction) and density bounded
on the low-density sides (east direction). The intensity gradient of the radio emission 
(see 610 MHz emission contours in Fig. \ref{fig6}b) supports that the ionized gas is escaping from the H{\sc ii} region towards the 
low-density ISM in the eastern direction. 
Based on radial velocity measurements of the  H$\alpha$ (i.e. -9.7 km s$^{-1}$) and 
CO ( i.e., -12.2 km s$^{-1}$) emissions, \citet{israel78} have also suggested that the ionized gas 
in the Sh2-112 region is probably undergoing a champagne flow.

Thus, the structure of the radio continuum map, dust extinction map, $^{13}$CO map, [S {\sc ii}]/[O {\sc iii}] ratio map, the spatial distribution of the young stars, 
and NIR to MIR images of the H{\sc ii} region indicate that the Sh2-112 is a blistered H{\sc ii} region in which
the ionized gas is possibly undergoing a champagne flow.

In Fig. \ref{morph}, we have shown the AKARI 160 $\micron$ image superimposed 
by Planck 353 GHz contours (left panel) and a simple geometry of the H{\sc ii} region (right panel). Based on the distribution of the cold dust emission and young stars, we suggest that the massive star possibly formed in the eastern edge of a cylindrical molecular cloud. It has ionized the surrounding molecular material and created the H{\sc ii} region. As the IF from the massive star reached the cloud surface, it broke the cloud boundary and quickly expanded to the eastern side (low-density ISM) while, in the western side of the ionizing star, the expansion is slow probably due to a higher-density molecular cloud. In the north-west boundary, an UC H{\sc ii} region is created 
by a massive member that is still embedded in the molecular cloud. Some of the YSO candidates observed 
toward the western periphery may be originated due to the interaction of the expanding H{\sc ii} region 
with the surrounding molecular cloud. 
However, with the present observations, it is difficult to disentangle the YSO candidates formed via triggered or spontaneous star formation scenario.  
\subsection{Star formation towards the south-west: possible case of triggering?}
To study the influence of the massive O8V star on the MSX1, we estimated the ionized gas pressure in the IBL and the internal pressure of the molecular condensation. 
A comparison of these  pressures can be used as a sensitive diagnostic for 
the induced star formation within the molecular cloud \citep{lefl94,morgan04}. 
If the internal pressure of the cloud is higher with respect to the IBL, then the photo-ionization induced shock will be stalled at the surface and hence it
is highly unlikely that any star formation activity observed within the cloud 
is due to triggering from the massive star. Conversely, if the cloud is under-pressured with respect to the IBL, then there is a high possibility that photo-ionization driven shocks are propagating or have already propagated through the molecular cloud and consequently, the star
formation within the clouds could have been induced due to radiative feedback effect of a massive star. 
Following \citet{thompson04}, the ionized gas pressure for the IBL is estimated using the equation:\\
\begin{equation}
	P_{i}/K_{B} = 2 {n_e}_{i} {T} 
\end{equation}
where $n_e$ is the electron density (see eq. 3), 
and $K_B$ is the Boltzman constant. We obtained a value of $\sim$ 8 $\times$ 10$^6$ cm$^{-3}$K for $P_i$/$K_B$. To estimate the internal pressure of the molecular condensation, we adopted the following relation:\\
\begin{equation}
	P_{int}/K_{B} \sim {\sigma}^2 {\rho}_{int}\\
\end{equation}
where $\rho$$_{int}$ is the internal density of the molecular condensation and 
$\sigma$ is the velocity dispersion. The value of $\sigma$$^2$ is estimated using the relation, $\sigma$$^2$= $\Delta$$v$$^2$ / (8 ln 2), where $\Delta$$v$ is the line width of the $^{13}$$CO$ (J = 3 - 2) line taken from \citet{urqu09}. Here, we assumed a number density of the molecular condensation, n(H$_2$), as 10$^3$ cm$^{-3}$. This yields a value of $P_{int}$/$K_B$ as $\sim$ 7.3 $\times$ 10$^5$ cm$^{-3}$ K. The main factors contributing to the uncertainties in the ionized gas pressure 
value are the electron density and the temperature of the ionized gas. The uncertainties involved in the calculation of electron density depend on the assumption of $\eta$ and the distance. The combined error of both of them may be about 40\% \citep{ortega13}. Similarly, the uncertainties involved with the value of internal 
molecular pressure include the uncertainties in the line width measurements and radius of the cloud. In case of a sample of bright-rimmed clouds, \citet{morgan04} estimated 
the uncertainties in the internal molecular pressures that were 
no more than a factor of 5.

A comparison between internal and ionized gas pressures indicates that the IBL is over-pressurized with respect to the molecular 
condensation by a factor of 10, suggesting that 
shocks are propagating /propagated through the molecular condensation and the young stars identified within it are likely triggered due to the massive star BD~+45 3216. 
Since the ionizing source is an O8V star, still on
the MS, hence it should have a MS lifetime of $<$ 5 Myr \citep{meynet94}.
This suggests that the H{\sc ii} region may still be under expansion and may be responsible for triggering next generation star formation. 
\section{Conclusions}
The complex nature of the H{\sc ii} region Sh2-112 has made it very intriguing and poorly studied H{\sc ii} region. Understanding the ongoing physical processes 
in such a complex region requires a thorough multiwavelength analysis.  
In the present work, we study the physical environment of a Galactic H{\sc ii} region Sh2-112 using multiwavelength observations. 
We use optical and NIR photometric and spectroscopic observations, as well as radio continuum observations. Our main 
findings are as follows:
\begin{itemize}
	\item The analysis of the optical spectra of the  bright source BD +45 3216 confirms that it is of O8 V type. 
	\item		Using various NIR/MIR CC and CM diagrams, we identified a total of 138 YSO candidates in the region. Out of these 8 are Class~{\sc i} objects, 59 are Class~{\sc ii} objects, and the remaining 71 are sources with IR excess ($H$ - $K$ $>$ 0.9) which could be candidate Class~{\sc ii}/Class~{\sc i} YSO candidates. The $H$/($H$ - $K$) CM diagram analysis shows that the majority of YSO candidates have masses less than 2 M$\odot$. 
		\item The NN surface density distribution and MST analyses of YSO candidates reveal the grouping of YSO candidates towards the 
western periphery. 
		\item The high-resolution 1280 MHz radio continuum map reveals a peak near the north-west periphery of the 
Sh2-112 region, which may be 
an UC H{\sc ii} region excited by a B0V-B0.5V star. Optical (H$\alpha$) and radio images show the presence of an IBL and molecular cloud emission at $^{13}$CO toward the south-west periphery. 
		This region also consists of a group of young stars. 
		\item We estimated the pressures in the IBL and molecular clump and found that the IBL is over-pressured with respect to the molecular condensation, suggesting that the photo-ionization driven shocks are propagating/propagated inside the molecular condensation and the star formation observed inside it may be due to the triggering by the massive star.

			The distribution of the dust emission, radio continuum 
		emission, [S {\sc ii}]/[O {\sc iii}] ratio map, and the surface density distribution of the YSO candidates suggest that Sh2-112 is a blister H{\sc ii} region which may have formed in a cylindrical molecular cloud.    
\end{itemize} 
\acknowledgments
We thank the anonymous reviewer for a critical reading of
the manuscript and constructive suggestions that greatly
improved the quality of the manuscript. NP acknowledges the financial support from the Department of Science \& Technology (DST),
INDIA, through INSPIRE faculty award~IFA-PH-36. SS and NP acknowledge the support of the DST, Government
of India, under project no. DST/INT/Thai/P-15/2019. DKO acknowledges the support of the Department of Atomic Energy, Government of India, under Project Identification No. RTI 4002. TB is supported by the National Key Research and Development Program of China (2017YFA0402702). TB also acknowledges the funding from the China Postdoctoral Science Foundation through grant 2018M631241, and PKU-Tokyo Partner fund. We thank the
staff of IAO, Hanle and CREST, Hosakote, that made these observations
possible. The facilities at IAO and CREST are operated by the Indian
Institute of Astrophysics, Bangalore. This publication makes use of data from the Two Micron All Sky Survey (a 
joint project of the University of Massachusetts and the Infrared Processing 
and Analysis Center/ California Institute of Technology, funded by the 
National Aeronautics and Space Administration and the National Science 
Foundation), archival data obtained with the {\it Spitzer Space Telescope} and {\it Wide Infrared Survey Explorer} 
(operated by the Jet Propulsion Laboratory, California Institute 
of Technology, under contract with the NASA.

\bibliographystyle{aasjournal}
\bibliography{ref}

\begin{thebibliography}{}
\expandafter\ifx\csname natexlab\endcsname\relax\def\natexlab#1{#1}\fi
\providecommand{\url}[1]{\href{#1}{#1}}
\providecommand{\dodoi}[1]{doi:~\href{http://doi.org/#1}{\nolinkurl{#1}}}
\providecommand{\doeprint}[1]{\href{http://ascl.net/#1}{\nolinkurl{http://ascl.net/#1}}}
\providecommand{\doarXiv}[1]{\href{https://arxiv.org/abs/#1}{\nolinkurl{https://arxiv.org/abs/#1}}}

\bibitem[{{Artigau} {et~al.}(2004){Artigau}, {Doyon}, {Vallee}, {Riopel}, \&
  {Nadeau}}]{arti04}
{Artigau}, E., {Doyon}, R., {Vallee}, P., {Riopel}, M., \& {Nadeau}, D. 2004,
  in Society of Photo-Optical Instrumentation Engineers (SPIE) Conference
  Series, Vol. 5492, Proceedings of the SPIE, ed. A.~F.~M. {Moorwood} \&
  M.~{Iye}, 1479--1486

\bibitem[{{Bailer-Jones} {et~al.}(2018){Bailer-Jones}, {Rybizki}, {Fouesneau},
  {Mantelet}, \& {Andrae}}]{jone18}
{Bailer-Jones}, C.~A.~L., {Rybizki}, J., {Fouesneau}, M., {Mantelet}, G., \&
  {Andrae}, R. 2018, AJ, 156, 58, \dodoi{10.3847/1538-3881/aacb21}

\bibitem[{{Baug} {et~al.}(2015){Baug}, {Ojha}, {Dewangan}, {Ninan}, {Bhatt},
  {Ghosh}, \& {Mallick}}]{baug15}
{Baug}, T., {Ojha}, D.~K., {Dewangan}, L.~K., {et~al.} 2015, MNRAS, 454, 4335,
  \dodoi{10.1093/mnras/stv2269}

\bibitem[{{Bertoldi}(1989)}]{bert89}
{Bertoldi}, F. 1989, ApJ, 346, 735, \dodoi{10.1086/168055}

\bibitem[{{Bessell} \& {Brett}(1988)}]{bessel88}
{Bessell}, M.~S., \& {Brett}, J.~M. 1988, PASP, 100, 1134,
  \dodoi{10.1086/132281}

\bibitem[{{Bisbas} {et~al.}(2009){Bisbas}, {W{\"u}nsch}, {Whitworth}, \&
  {Hubber}}]{bisbas09}
{Bisbas}, T.~G., {W{\"u}nsch}, R., {Whitworth}, A.~P., \& {Hubber}, D.~A. 2009,
  A\&A, 497, 649, \dodoi{10.1051/0004-6361/200811522}

\bibitem[{{Buckle} {et~al.}(2009){Buckle}, {Hills}, {Smith}, {Dent}, {Bell},
  {Curtis}, {Dace}, {Gibson}, {Graves}, {Leech}, {Richer}, {Williamson},
  {Withington}, {Yassin}, {Bennett}, {Hastings}, {Laidlaw}, {Lightfoot},
  {Burgess}, {Dewdney}, {Hovey}, {Willis}, {Redman}, {Wooff}, {Berry},
  {Cavanagh}, {Davis}, {Dempsey}, {Friberg}, {Jenness}, {Kackley}, {Rees},
  {Tilanus}, {Walther}, {Zwart}, {Klapwijk}, {Kroug}, \& {Zijlstra}}]{buck09}
{Buckle}, J.~V., {Hills}, R.~E., {Smith}, H., {et~al.} 2009, MNRAS, 399, 1026,
  \dodoi{10.1111/j.1365-2966.2009.15347.x}

\bibitem[{{Cartwright} \& {Whitworth}(2004)}]{cartwright04}
{Cartwright}, A., \& {Whitworth}, A.~P. 2004, MNRAS, 348, 589,
  \dodoi{10.1111/j.1365-2966.2004.07360.x}

\bibitem[{{Chauhan} {et~al.}(2009){Chauhan}, {Pandey}, {Ogura}, {Ojha},
  {Bhatt}, {Ghosh}, \& {Rawat}}]{chauhan09}
{Chauhan}, N., {Pandey}, A.~K., {Ogura}, K., {et~al.} 2009, MNRAS, 396, 964,
  \dodoi{10.1111/j.1365-2966.2009.14756.x}

\bibitem[{{Chavarr{\'\i}a} {et~al.}(2014){Chavarr{\'\i}a}, {Allen}, {Brunt},
  {Hora}, {Muench}, \& {Fazio}}]{chav14}
{Chavarr{\'\i}a}, L., {Allen}, L., {Brunt}, C., {et~al.} 2014, MNRAS, 439,
  3719, \dodoi{10.1093/mnras/stu224}

\bibitem[{{Chrysostomou} {et~al.}(1992){Chrysostomou}, {Brand}, {Burton}, \&
  {Moorhouse}}]{chrys92}
{Chrysostomou}, A., {Brand}, P. W.~J.~L., {Burton}, M.~G., \& {Moorhouse}, A.
  1992, MNRAS, 256, 528, \dodoi{10.1093/mnras/256.3.528}

\bibitem[{{Cohen} {et~al.}(1981){Cohen}, {Frogel}, {Persson}, \&
  {Elias}}]{cohen81}
{Cohen}, J.~G., {Frogel}, J.~A., {Persson}, S.~E., \& {Elias}, J.~H. 1981, ApJ,
  249, 481, \dodoi{10.1086/159308}

\bibitem[{{Cooper} {et~al.}(2013){Cooper}, {Lumsden}, {Oudmaijer}, {Hoare},
  {Clarke}, {Urquhart}, {Mottram}, {Moore}, \& {Davies}}]{cooper13}
{Cooper}, H.~D.~B., {Lumsden}, S.~L., {Oudmaijer}, R.~D., {et~al.} 2013, MNRAS,
  430, 1125, \dodoi{10.1093/mnras/sts681}

\bibitem[{{Cutri} {et~al.}(2003){Cutri}, {Skrutskie}, {van Dyk}, {Beichman},
  {Carpenter}, {Chester}, {Cambresy}, {Evans}, {Fowler}, {Gizis}, {Howard},
  {Huchra}, {Jarrett}, {Kopan}, {Kirkpatrick}, {Light}, {Marsh}, {McCallon},
  {Schneider}, {Stiening}, {Sykes}, {Weinberg}, {Wheaton}, {Wheelock}, \&
  {Zacarias}}]{cutr03}
{Cutri}, R.~M., {Skrutskie}, M.~F., {van Dyk}, S., {et~al.} 2003, {2MASS All
  Sky Catalog of point sources.}

\bibitem[{{Cutri~{et al.}}(2014)}]{cutr14}
{Cutri~{et al.}} 2014, VizieR Online Data Catalog, II/328

\bibitem[{{Dale} {et~al.}(2007){Dale}, {Bonnell}, \& {Whitworth}}]{dale07}
{Dale}, J.~E., {Bonnell}, I.~A., \& {Whitworth}, A.~P. 2007, MNRAS, 375, 1291,
  \dodoi{10.1111/j.1365-2966.2006.11368.x}

\bibitem[{{Deharveng} {et~al.}(2005){Deharveng}, {Zavagno}, \&
  {Caplan}}]{deha05}
{Deharveng}, L., {Zavagno}, A., \& {Caplan}, J. 2005, A\&A, 433, 565,
  \dodoi{10.1051/0004-6361:20041946}

\bibitem[{{Deharveng} {et~al.}(2010){Deharveng}, {Schuller}, {Anderson},
  {Zavagno}, {Wyrowski}, {Menten}, {Bronfman}, {Testi}, {Walmsley}, \&
  {Wienen}}]{deharveng10}
{Deharveng}, L., {Schuller}, F., {Anderson}, L.~D., {et~al.} 2010, A\&A, 523,
  A6, \dodoi{10.1051/0004-6361/201014422}

\bibitem[{{Dewangan} {et~al.}(2018){Dewangan}, {Baug}, {Ojha}, \&
  {Ghosh}}]{dewa18}
{Dewangan}, L.~K., {Baug}, T., {Ojha}, D.~K., \& {Ghosh}, S.~K. 2018, ApJ, 869,
  30, \dodoi{10.3847/1538-4357/aae9db}

\bibitem[{{Dewangan} {et~al.}(2015){Dewangan}, {Luna}, {Ojha}, {Anand arao},
  {Mallick}, \& {Mayya}}]{dewa15}
{Dewangan}, L.~K., {Luna}, A., {Ojha}, D.~K., {et~al.} 2015, ApJ, 811, 79,
  \dodoi{10.1088/0004-637X/811/2/79}

\bibitem[{{Dewangan} {et~al.}(2019){Dewangan}, {Sano}, {Enokiya}, {Tachihara},
  {Fukui}, \& {Ojha}}]{dewangan19}
{Dewangan}, L.~K., {Sano}, H., {Enokiya}, R., {et~al.} 2019, ApJ, 878, 26,
  \dodoi{10.3847/1538-4357/ab1cba}

\bibitem[{{Dobashi} {et~al.}(1994){Dobashi}, {Bernard}, {Yonekura}, \&
  {Fukui}}]{dobashi94}
{Dobashi}, K., {Bernard}, J.-P., {Yonekura}, Y., \& {Fukui}, Y. 1994, ApJS, 95,
  419, \dodoi{10.1086/192106}

\bibitem[{{Downes} \& {Rinehart}(1966)}]{down66}
{Downes}, D., \& {Rinehart}, R. 1966, ApJ, 144, 937, \dodoi{10.1086/148691}

\bibitem[{{Duronea} {et~al.}(2012){Duronea}, {Vasquez}, {Cappa}, {Corti}, \&
  {Arnal}}]{duronea12}
{Duronea}, N.~U., {Vasquez}, J., {Cappa}, C.~E., {Corti}, M., \& {Arnal}, E.~M.
  2012, A\&A, 537, A149, \dodoi{10.1051/0004-6361/201117958}

\bibitem[{{Dyson} \& {Williams}(1980)}]{dyson80}
{Dyson}, J.~E., \& {Williams}, D.~A. 1980, {Physics of the interstellar medium}

\bibitem[{{Elmegreen}(1998)}]{elme98}
{Elmegreen}, B.~G. 1998, Astronomical Society of the Pacific Conference Series,
  Vol. 148, {Observations and Theory of Dynamical Triggers for Star Formation},
  ed. C.~E. {Woodward}, J.~M. {Shull}, \& J.~{Thronson}, Harley~A., 150

\bibitem[{{Elmegreen} \& {Lada}(1977)}]{elme77}
{Elmegreen}, B.~G., \& {Lada}, C.~J. 1977, ApJ, 214, 725,
  \dodoi{10.1086/155302}

\bibitem[{{Flaherty} {et~al.}(2007){Flaherty}, {Pipher}, {Megeath}, {Winston},
  {Gutermuth}, {Muzerolle}, {Allen}, \& {Fazio}}]{flaherty07}
{Flaherty}, K.~M., {Pipher}, J.~L., {Megeath}, S.~T., {et~al.} 2007, ApJ, 663,
  1069, \dodoi{10.1086/518411}

\bibitem[{{Gaia Collaboration} {et~al.}(2018){Gaia Collaboration}, {Brown},
  {Vallenari}, {Prusti}, {de Bruijne}, {Babusiaux}, {Bailer-Jones}, {Biermann},
  {Evans}, {Eyer}, {Jansen}, {Jordi}, {Klioner}, {Lammers}, {Lindegren},
  {Luri}, {Mignard}, {Panem}, {Pourbaix}, {Randich}, {Sartoretti}, {Siddiqui},
  {Soubiran}, {van Leeuwen}, {Walton}, {Arenou}, {Bastian}, {Cropper},
  {Drimmel}, {Katz}, {Lattanzi}, {Bakker}, {Cacciari}, {Casta{\~n}eda},
  {Chaoul}, {Cheek}, {De Angeli}, {Fabricius}, {Guerra}, {Holl}, {Masana},
  {Messineo}, {Mowlavi}, {Nienartowicz}, {Panuzzo}, {Portell}, {Riello},
  {Seabroke}, {Tanga}, {Th{\'e}venin}, {Gracia-Abril}, {Comoretto},
  {Garcia-Reinaldos}, {Teyssier}, {Altmann}, {Andrae}, {Audard},
  {Bellas-Velidis}, {Benson}, {Berthier}, {Blomme}, {Burgess}, {Busso},
  {Carry}, {Cellino}, {Clementini}, {Clotet}, {Creevey}, {Davidson}, {De
  Ridder}, {Delchambre}, {Dell'Oro}, {Ducourant},
  {Fern{\'a}ndez-Hern{\'a}ndez}, {Fouesneau}, {Fr{\'e}mat}, {Galluccio},
  {Garc{\'\i}a-Torres}, {Gonz{\'a}lez-N{\'u}{\~n}ez}, {Gonz{\'a}lez-Vidal},
  {Gosset}, {Guy}, {Halbwachs}, {Hambly}, {Harrison}, {Hern{\'a}ndez},
  {Hestroffer}, {Hodgkin}, {Hutton}, {Jasniewicz}, {Jean-Antoine-Piccolo},
  {Jordan}, {Korn}, {Krone-Martins}, {Lanzafame}, {Lebzelter}, {L{\"o}ffler},
  {Manteiga}, {Marrese}, {Mart{\'\i}n-Fleitas}, {Moitinho}, {Mora}, {Muinonen},
  {Osinde}, {Pancino}, {Pauwels}, {Petit}, {Recio-Blanco}, {Richards},
  {Rimoldini}, {Robin}, {Sarro}, {Siopis}, {Smith}, {Sozzetti}, {S{\"u}veges},
  {Torra}, {van Reeven}, {Abbas}, {Abreu Aramburu}, {Accart}, {Aerts},
  {Altavilla}, {{\'A}lvarez}, {Alvarez}, {Alves}, {Anderson}, {Andrei},
  {Anglada Varela}, {Antiche}, {Antoja}, {Arcay}, {Astraatmadja}, {Bach},
  {Baker}, {Balaguer-N{\'u}{\~n}ez}, {Balm}, {Barache}, {Barata}, {Barbato},
  {Barblan}, {Barklem}, {Barrado}, {Barros}, {Barstow}, {Bartholom{\'e}
  Mu{\~n}oz}, {Bassilana}, {Becciani}, {Bellazzini}, {Berihuete}, {Bertone},
  {Bianchi}, {Bienaym{\'e}}, {Blanco-Cuaresma}, {Boch}, {Boeche}, {Bombrun},
  {Borrachero}, {Bossini}, {Bouquillon}, {Bourda}, {Bragaglia}, {Bramante},
  {Breddels}, {Bressan}, {Brouillet}, {Br{\"u}semeister}, {Brugaletta},
  {Bucciarelli}, {Burlacu}, {Busonero}, {Butkevich}, {Buzzi}, {Caffau},
  {Cancelliere}, {Cannizzaro}, {Cantat-Gaudin}, {Carballo}, {Carlucci},
  {Carrasco}, {Casamiquela}, {Castellani}, {Castro-Ginard}, {Charlot},
  {Chemin}, {Chiavassa}, {Cocozza}, {Costigan}, {Cowell}, {Crifo}, {Crosta},
  {Crowley}, {Cuypers}, {Dafonte}, {Damerdji}, {Dapergolas}, {David}, {David},
  {de Laverny}, {De Luise}, {De March}, {de Martino}, {de Souza}, {de Torres},
  {Debosscher}, {del Pozo}, {Delbo}, {Delgado}, {Delgado}, {Di Matteo},
  {Diakite}, {Diener}, {Distefano}, {Dolding}, {Drazinos}, {Dur{\'a}n},
  {Edvardsson}, {Enke}, {Eriksson}, {Esquej}, {Eynard Bontemps}, {Fabre},
  {Fabrizio}, {Faigler}, {Falc{\~a}o}, {Farr{\`a}s Casas}, {Federici},
  {Fedorets}, {Fernique}, {Figueras}, {Filippi}, {Findeisen}, {Fonti},
  {Fraile}, {Fraser}, {Fr{\'e}zouls}, {Gai}, {Galleti}, {Garabato},
  {Garc{\'\i}a-Sedano}, {Garofalo}, {Garralda}, {Gavel}, {Gavras}, {Gerssen},
  {Geyer}, {Giacobbe}, {Gilmore}, {Girona}, {Giuffrida}, {Glass}, {Gomes},
  {Granvik}, {Gueguen}, {Guerrier}, {Guiraud}, {Guti{\'e}rrez-S{\'a}nchez},
  {Haigron}, {Hatzidimitriou}, {Hauser}, {Haywood}, {Heiter}, {Helmi}, {Heu},
  {Hilger}, {Hobbs}, {Hofmann}, {Holland}, {Huckle}, {Hypki}, {Icardi},
  {Jan{\ss}en}, {Jevardat de Fombelle}, {Jonker}, {Juh{\'a}sz}, {Julbe},
  {Karampelas}, {Kewley}, {Klar}, {Kochoska}, {Kohley}, {Kolenberg},
  {Kontizas}, {Kontizas}, {Koposov}, {Kordopatis}, {Kostrzewa-Rutkowska},
  {Koubsky}, {Lambert}, {Lanza}, {Lasne}, {Lavigne}, {Le Fustec}, {Le
  Poncin-Lafitte}, {Lebreton}, {Leccia}, {Leclerc}, {Lecoeur-Taibi},
  {Lenhardt}, {Leroux}, {Liao}, {Licata}, {Lindstr{\o}m}, {Lister}, {Livanou},
  {Lobel}, {L{\'o}pez}, {Managau}, {Mann}, {Mantelet}, {Marchal}, {Marchant},
  {Marconi}, {Marinoni}, {Marschalk{\'o}}, {Marshall}, {Martino}, {Marton},
  {Mary}, {Massari}, {Matijevi{\v{c}}}, {Mazeh}, {McMillan}, {Messina},
  {Michalik}, {Millar}, {Molina}, {Molinaro}, {Moln{\'a}r}, {Montegriffo},
  {Mor}, {Morbidelli}, {Morel}, {Morris}, {Mulone}, {Muraveva}, {Musella},
  {Nelemans}, {Nicastro}, {Noval}, {O'Mullane}, {Ord{\'e}novic},
  {Ord{\'o}{\~n}ez-Blanco}, {Osborne}, {Pagani}, {Pagano}, {Pailler},
  {Palacin}, {Palaversa}, {Panahi}, {Pawlak}, {Piersimoni}, {Pineau}, {Plachy},
  {Plum}, {Poggio}, {Poujoulet}, {Pr{\v{s}}a}, {Pulone}, {Racero}, {Ragaini},
  {Rambaux}, {Ramos-Lerate}, {Regibo}, {Reyl{\'e}}, {Riclet}, {Ripepi}, {Riva},
  {Rivard}, {Rixon}, {Roegiers}, {Roelens}, {Romero-G{\'o}mez}, {Rowell},
  {Royer}, {Ruiz-Dern}, {Sadowski}, {Sagrist{\`a} Sell{\'e}s}, {Sahlmann},
  {Salgado}, {Salguero}, {Sanna}, {Santana-Ros}, {Sarasso}, {Savietto},
  {Schultheis}, {Sciacca}, {Segol}, {Segovia}, {S{\'e}gransan}, {Shih},
  {Siltala}, {Silva}, {Smart}, {Smith}, {Solano}, {Solitro}, {Sordo}, {Soria
  Nieto}, {Souchay}, {Spagna}, {Spoto}, {Stampa}, {Steele},
  {Steidelm{\"u}ller}, {Stephenson}, {Stoev}, {Suess}, {Surdej}, {Szabados},
  {Szegedi-Elek}, {Tapiador}, {Taris}, {Tauran}, {Taylor}, {Teixeira},
  {Terrett}, {Teyssand ier}, {Thuillot}, {Titarenko}, {Torra Clotet}, {Turon},
  {Ulla}, {Utrilla}, {Uzzi}, {Vaillant}, {Valentini}, {Valette}, {van Elteren},
  {Van Hemelryck}, {van Leeuwen}, {Vaschetto}, {Vecchiato}, {Veljanoski},
  {Viala}, {Vicente}, {Vogt}, {von Essen}, {Voss}, {Votruba}, {Voutsinas},
  {Walmsley}, {Weiler}, {Wertz}, {Wevers}, {Wyrzykowski}, {Yoldas},
  {{\v{Z}}erjal}, {Ziaeepour}, {Zorec}, {Zschocke}, {Zucker}, {Zurbach}, \&
  {Zwitter}}]{gaia18}
{Gaia Collaboration}, {Brown}, A.~G.~A., {Vallenari}, A., {et~al.} 2018, A\&A,
  616, A1, \dodoi{10.1051/0004-6361/201833051}

\bibitem[{{Girardi} {et~al.}(2002){Girardi}, {Bertelli}, {Bressan}, {Chiosi},
  {Groenewegen}, {Marigo}, {Salasnich}, \& {Weiss}}]{girardi02}
{Girardi}, L., {Bertelli}, G., {Bressan}, A., {et~al.} 2002, A\&A, 391, 195,
  \dodoi{10.1051/0004-6361:20020612}

\bibitem[{{Goicoechea} {et~al.}(2018){Goicoechea}, {Cuadrado}, {Pety},
  {Aguado}, {Black}, {Bron}, {Cernicharo}, {Chapillon}, {Fuente}, {Gerin},
  {Joblin}, {Roncero}, \& {Tercero}}]{goico18}
{Goicoechea}, J.~R., {Cuadrado}, S., {Pety}, J., {et~al.} 2018, in IAU
  Symposium, Vol. 332, IAU Symposium, ed. M.~{Cunningham}, T.~{Millar}, \&
  Y.~{Aikawa}, 210--217

\bibitem[{{Gutermuth} {et~al.}(2009){Gutermuth}, {Megeath}, {Myers}, {Allen},
  {Pipher}, \& {Fazio}}]{guter09}
{Gutermuth}, R.~A., {Megeath}, S.~T., {Myers}, P.~C., {et~al.} 2009, ApJS, 184,
  18, \dodoi{10.1088/0067-0049/184/1/18}

\bibitem[{{Henney} {et~al.}(2005){Henney}, {Arthur}, \&
  {Garc{\'\i}a-D{\'\i}az}}]{henney05}
{Henney}, W.~J., {Arthur}, S.~J., \& {Garc{\'\i}a-D{\'\i}az}, M.~T. 2005, ApJ,
  627, 813, \dodoi{10.1086/430593}

\bibitem[{{Hern{\'a}ndez} {et~al.}(2005){Hern{\'a}ndez}, {Calvet}, {Hartmann},
  {Brice{\~n}o}, {Sicilia-Aguilar}, \& {Berlind}}]{hern05}
{Hern{\'a}ndez}, J., {Calvet}, N., {Hartmann}, L., {et~al.} 2005, AJ, 129, 856,
  \dodoi{10.1086/426918}

\bibitem[{{Hosokawa} \& {Inutsuka}(2006)}]{hoso06}
{Hosokawa}, T., \& {Inutsuka}, S.-i. 2006, ApJ, 646, 240,
  \dodoi{10.1086/504789}

\bibitem[{{Hunter} \& {Massey}(1990)}]{hunt90}
{Hunter}, D.~A., \& {Massey}, P. 1990, AJ, 99, 846, \dodoi{10.1086/115378}

\bibitem[{{Israel}(1978)}]{israel78}
{Israel}, F.~P. 1978, A\&A, 70, 769

\bibitem[{{Jacoby} {et~al.}(1984){Jacoby}, {Hunter}, \& {Christian}}]{jaco84}
{Jacoby}, G.~H., {Hunter}, D.~A., \& {Christian}, C.~A. 1984, ApJS, 56, 257,
  \dodoi{10.1086/190983}

\bibitem[{{Kn{\"o}dlseder}(2000)}]{knod00}
{Kn{\"o}dlseder}, J. 2000, A\&A, 360, 539.
\newblock \doarXiv{astro-ph/0007442}

\bibitem[{{Koenig} \& {Leisawitz}(2014)}]{koenig14}
{Koenig}, X.~P., \& {Leisawitz}, D.~T. 2014, ApJ, 791, 131,
  \dodoi{10.1088/0004-637X/791/2/131}

\bibitem[{{Kruijssen} {et~al.}(2019){Kruijssen}, {Schruba}, {Chevance},
  {Longmore}, {Hygate}, {Haydon}, {McLeod}, {Dalcanton}, {Tacconi}, \& {van
  Dishoeck}}]{krui19}
{Kruijssen}, J.~M.~D., {Schruba}, A., {Chevance}, M., {et~al.} 2019, Nature,
  569, 519, \dodoi{10.1038/s41586-019-1194-3}

\bibitem[{{Krumholz} \& {Matzner}(2009)}]{krumholz09}
{Krumholz}, M.~R., \& {Matzner}, C.~D. 2009, ApJ, 703, 1352,
  \dodoi{10.1088/0004-637X/703/2/1352}

\bibitem[{{Kwan}(1997)}]{kwan97}
{Kwan}, J. 1997, ApJ, 489, 284, \dodoi{10.1086/304773}

\bibitem[{{Lada} \& {Adams}(1992)}]{lada92}
{Lada}, C.~J., \& {Adams}, F.~C. 1992, ApJ, 393, 278, \dodoi{10.1086/171505}

\bibitem[{{Lada} {et~al.}(1994){Lada}, {Lada}, {Clemens}, \& {Bally}}]{lada94}
{Lada}, C.~J., {Lada}, E.~A., {Clemens}, D.~P., \& {Bally}, J. 1994, ApJ, 429,
  694, \dodoi{10.1086/174354}

\bibitem[{{Lahulla}(1985)}]{lahu85}
{Lahulla}, J.~F. 1985, A\&ASS, 61, 537

\bibitem[{{Lawrence} {et~al.}(2007){Lawrence}, {Warren}, {Almaini}, {Edge},
  {Hambly}, {Jameson}, {Lucas}, {Casali}, {Adamson}, {Dye}, {Emerson},
  {Foucaud}, {Hewett}, {Hirst}, {Hodgkin}, {Irwin}, {Lodieu}, {McMahon},
  {Simpson}, {Smail}, {Mortlock}, \& {Folger}}]{lawr07}
{Lawrence}, A., {Warren}, S.~J., {Almaini}, O., {et~al.} 2007, MNRAS, 379,
  1599, \dodoi{10.1111/j.1365-2966.2007.12040.x}

\bibitem[{{Lefloch} \& {Lazareff}(1994)}]{lefl94}
{Lefloch}, B., \& {Lazareff}, B. 1994, A\&A, 289, 559

\bibitem[{{Lefloch} \& {Lazareff}(1995)}]{lefloch95}
---. 1995, A\&A, 301, 522

\bibitem[{{Lefloch} {et~al.}(1997){Lefloch}, {Lazareff}, \&
  {Castets}}]{lefloch97}
{Lefloch}, B., {Lazareff}, B., \& {Castets}, A. 1997, A\&A, 324, 249

\bibitem[{{Lucas} {et~al.}(2008){Lucas}, {Hoare}, {Longmore}, {Schr{\"o}der},
  {Davis}, {Adamson}, {Band yopadhyay}, {de Grijs}, {Smith}, {Gosling},
  {Mitchison}, {G{\'a}sp{\'a}r}, {Coe}, {Tamura}, {Parker}, {Irwin}, {Hambly},
  {Bryant}, {Collins}, {Cross}, {Evans}, {Gonzalez-Solares}, {Hodgkin},
  {Lewis}, {Read}, {Riello}, {Sutorius}, {Lawrence}, {Drew}, {Dye}, \&
  {Thompson}}]{lucas08}
{Lucas}, P.~W., {Hoare}, M.~G., {Longmore}, A., {et~al.} 2008, MNRAS, 391, 136,
  \dodoi{10.1111/j.1365-2966.2008.13924.x}

\bibitem[{{Luri} {et~al.}(2018){Luri}, {Brown}, {Sarro}, {Arenou},
  {Bailer-Jones}, {Castro-Ginard}, {de Bruijne}, {Prusti}, {Babusiaux}, \&
  {Delgado}}]{luri18}
{Luri}, X., {Brown}, A.~G.~A., {Sarro}, L.~M., {et~al.} 2018, A\&A, 616, A9,
  \dodoi{10.1051/0004-6361/201832964}

\bibitem[{{Maheswar} {et~al.}(2007){Maheswar}, {Sharma}, {Biman}, {Pand ey}, \&
  {Bhatt}}]{2007MNRAS.379.1237M}
{Maheswar}, G., {Sharma}, S., {Biman}, J.~M., {Pand ey}, A.~K., \& {Bhatt},
  H.~C. 2007, \mnras, 379, 1237, \dodoi{10.1111/j.1365-2966.2007.12020.x}

\bibitem[{{Mallick} {et~al.}(2013){Mallick}, {Kumar}, {Ojha}, {Bachiller},
  {Samal}, \& {Pirogov}}]{mall13}
{Mallick}, K.~K., {Kumar}, M.~S.~N., {Ojha}, D.~K., {et~al.} 2013, ApJ, 779,
  113, \dodoi{10.1088/0004-637X/779/2/113}

\bibitem[{{Mallick} {et~al.}(2012){Mallick}, {Ojha}, {Samal}, {Pand ey},
  {Bhatt}, {Ghosh}, {Dewangan}, \& {Tamura}}]{mall12}
{Mallick}, K.~K., {Ojha}, D.~K., {Samal}, M.~R., {et~al.} 2012, ApJ, 759, 48,
  \dodoi{10.1088/0004-637X/759/1/48}

\bibitem[{{Martins} {et~al.}(2005){Martins}, {Schaerer}, \&
  {Hillier}}]{martins05}
{Martins}, F., {Schaerer}, D., \& {Hillier}, D.~J. 2005, A\&A, 436, 1049,
  \dodoi{10.1051/0004-6361:20042386}

\bibitem[{{Maud} {et~al.}(2015{\natexlab{a}}){Maud}, {Lumsden}, {Moore},
  {Mottram}, {Urquhart}, \& {Cicchini}}]{maud15a}
{Maud}, L.~T., {Lumsden}, S.~L., {Moore}, T.~J.~T., {et~al.}
  2015{\natexlab{a}}, MNRAS, 452, 637, \dodoi{10.1093/mnras/stv1334}

\bibitem[{{Maud} {et~al.}(2015{\natexlab{b}}){Maud}, {Moore}, {Lumsden},
  {Mottram}, {Urquhart}, \& {Hoare}}]{maud15b}
{Maud}, L.~T., {Moore}, T.~J.~T., {Lumsden}, S.~L., {et~al.}
  2015{\natexlab{b}}, MNRAS, 453, 645, \dodoi{10.1093/mnras/stv1635}

\bibitem[{{Meyer} {et~al.}(1997){Meyer}, {Calvet}, \& {Hillenbrand}}]{meyer97}
{Meyer}, M.~R., {Calvet}, N., \& {Hillenbrand}, L.~A. 1997, AJ, 114, 288,
  \dodoi{10.1086/118474}

\bibitem[{{Meynet} {et~al.}(1994){Meynet}, {Maeder}, {Schaller}, {Schaerer}, \&
  {Charbonnel}}]{meynet94}
{Meynet}, G., {Maeder}, A., {Schaller}, G., {Schaerer}, D., \& {Charbonnel}, C.
  1994, A\&ASS, 103, 97

\bibitem[{{Morgan} {et~al.}(2004){Morgan}, {Thompson}, {Urquhart}, {White}, \&
  {Miao}}]{morgan04}
{Morgan}, L.~K., {Thompson}, M.~A., {Urquhart}, J.~S., {White}, G.~J., \&
  {Miao}, J. 2004, A\&A, 426, 535, \dodoi{10.1051/0004-6361:20040226}

\bibitem[{{Motte} {et~al.}(2018){Motte}, {Bontemps}, \& {Louvet}}]{mott18}
{Motte}, F., {Bontemps}, S., \& {Louvet}, F. 2018, ARA\&A, 56, 41,
  \dodoi{10.1146/annurev-astro-091916-055235}

\bibitem[{{Ojha} {et~al.}(2004{\natexlab{a}}){Ojha}, {Ghosh}, {Kulkarni},
  {Testi}, {Verma}, \& {Vig}}]{ojha04a}
{Ojha}, D.~K., {Ghosh}, S.~K., {Kulkarni}, V.~K., {et~al.} 2004{\natexlab{a}},
  A\&A, 415, 1039, \dodoi{10.1051/0004-6361:20034312}

\bibitem[{{Ojha} {et~al.}(2004{\natexlab{b}}){Ojha}, {Tamura}, {Nakajima},
  {Fukagawa}, {Sugitani}, {Nagashima}, {Nagayama}, {Nagata}, {Sato}, {Vig},
  {Ghosh}, {Pickles}, {Momose}, \& {Ogura}}]{ojha04b}
{Ojha}, D.~K., {Tamura}, M., {Nakajima}, Y., {et~al.} 2004{\natexlab{b}}, ApJ,
  616, 1042, \dodoi{10.1086/425068}

\bibitem[{{Omar} {et~al.}(2002){Omar}, {Chengalur}, \& {Anish Roshi}}]{omar02}
{Omar}, A., {Chengalur}, J.~N., \& {Anish Roshi}, D. 2002, A\&A, 395, 227,
  \dodoi{10.1051/0004-6361:20021302}

\bibitem[{{Ortega} {et~al.}(2013){Ortega}, {Paron}, {Giacani}, {Rubio}, \&
  {Dubner}}]{ortega13}
{Ortega}, M.~E., {Paron}, S., {Giacani}, E., {Rubio}, M., \& {Dubner}, G. 2013,
  A\&A, 556, A105, \dodoi{10.1051/0004-6361/201321808}

\bibitem[{{Paladini} {et~al.}(2012){Paladini}, {Umana}, {Veneziani},
  {Noriega-Crespo}, {Anderson}, {Piacentini}, {Pinheiro Gon{\c{c}}alves},
  {Paradis}, {Tibbs}, {Bernard}, \& {Natoli}}]{paladini12}
{Paladini}, R., {Umana}, G., {Veneziani}, M., {et~al.} 2012, ApJ, 760, 149,
  \dodoi{10.1088/0004-637X/760/2/149}

\bibitem[{{Panagia}(1973)}]{panagia73}
{Panagia}, N. 1973, AJ, 78, 929, \dodoi{10.1086/111498}

\bibitem[{{Pandey} {et~al.}(2020){Pandey}, {Sharma}, {Panwar}, {Dewangan},
  {Ojha}, {Bisen}, {Sinha}, {Ghosh}, \& {Pandey}}]{pandey20}
{Pandey}, R., {Sharma}, S., {Panwar}, N., {et~al.} 2020, \apj, 891, 81,
  \dodoi{10.3847/1538-4357/ab6dc7}

\bibitem[{{Panwar} {et~al.}(2014){Panwar}, {Chen}, {Pandey}, {Samal}, {Ogura},
  {Ojha}, {Jose}, \& {Bhatt}}]{panwar14}
{Panwar}, N., {Chen}, W.~P., {Pandey}, A.~K., {et~al.} 2014, MNRAS, 443, 1614,
  \dodoi{10.1093/mnras/stu1244}

\bibitem[{{Panwar} {et~al.}(2017){Panwar}, {Samal}, {Pandey}, {Jose}, {Chen},
  {Ojha}, {Ogura}, {Singh}, \& {Yadav}}]{panwar2017}
{Panwar}, N., {Samal}, M.~R., {Pandey}, A.~K., {et~al.} 2017, MNRAS, 468, 2684,
  \dodoi{10.1093/mnras/stx616}

\bibitem[{{Pellegrini} {et~al.}(2012){Pellegrini}, {Oey}, {Winkler}, {Points},
  {Smith}, {Jaskot}, \& {Zastrow}}]{pelle12}
{Pellegrini}, E.~W., {Oey}, M.~S., {Winkler}, P.~F., {et~al.} 2012, ApJ, 755,
  40, \dodoi{10.1088/0004-637X/755/1/40}

\bibitem[{{Pomar{\`e}s} {et~al.}(2009){Pomar{\`e}s}, {Zavagno}, {Deharveng},
  {Cunningham}, {Jones}, {Kurtz}, {Russeil}, {Caplan}, \&
  {Comer{\'o}n}}]{poma09}
{Pomar{\`e}s}, M., {Zavagno}, A., {Deharveng}, L., {et~al.} 2009, A\&A, 494,
  987, \dodoi{10.1051/0004-6361:200811050}

\bibitem[{{Samal} {et~al.}(2007){Samal}, {Pandey}, {Ojha}, {Ghosh}, {Kulkarni},
  \& {Bhatt}}]{sama07}
{Samal}, M.~R., {Pandey}, A.~K., {Ojha}, D.~K., {et~al.} 2007, ApJ, 671, 555,
  \dodoi{10.1086/522941}

\bibitem[{{Schmeja} \& {Klessen}(2006)}]{schmeja06}
{Schmeja}, S., \& {Klessen}, R.~S. 2006, A\&A, 449, 151,
  \dodoi{10.1051/0004-6361:20054464}

\bibitem[{{Schneider} {et~al.}(2006){Schneider}, {Bontemps}, {Simon}, {Jakob},
  {Motte}, {Miller}, {Kramer}, \& {Stutzki}}]{schn06}
{Schneider}, N., {Bontemps}, S., {Simon}, R., {et~al.} 2006, A\&A, 458, 855,
  \dodoi{10.1051/0004-6361:20065088}

\bibitem[{{Sharma} {et~al.}(2017){Sharma}, {Pandey}, {Ojha}, {Bhatt}, {Ogura},
  {Kobayashi}, {Yadav}, \& {Pandey}}]{sharma17}
{Sharma}, S., {Pandey}, A.~K., {Ojha}, D.~K., {et~al.} 2017, MNRAS, 467, 2943,
  \dodoi{10.1093/mnras/stx014}

\bibitem[{{Sharma} {et~al.}(2016){Sharma}, {Pandey}, {Borissova}, {Ojha},
  {Ivanov}, {Ogura}, {Kobayashi}, {Kurtev}, {Gopinathan}, \& {Kesh
  Yadav}}]{sharma16}
{Sharma}, S., {Pandey}, A.~K., {Borissova}, J., {et~al.} 2016, AJ, 151, 126,
  \dodoi{10.3847/0004-6256/151/5/126}

\bibitem[{{Sharpless}(1959)}]{sharpless59}
{Sharpless}, S. 1959, ApJS, 4, 257, \dodoi{10.1086/190049}

\bibitem[{{Siess} {et~al.}(2000){Siess}, {Dufour}, \& {Forestini}}]{siess2000}
{Siess}, L., {Dufour}, E., \& {Forestini}, M. 2000, A\&A, 358, 593.
\newblock \doarXiv{astro-ph/0003477}

\bibitem[{{Stalin} {et~al.}(2008){Stalin}, {Hegde}, {Sahu}, {Parihar},
  {Anupama}, {Bhatt}, \& {Prabhu}}]{stalin08}
{Stalin}, C.~S., {Hegde}, M., {Sahu}, D.~K., {et~al.} 2008, Bulletin of the
  Astronomical Society of India, 36, 111.
\newblock \doarXiv{0809.1745}

\bibitem[{{Stone}(1977)}]{ston77}
{Stone}, R.~P.~S. 1977, ApJ, 218, 767, \dodoi{10.1086/155732}

\bibitem[{{Tan} {et~al.}(2014){Tan}, {Beltr{\'a}n}, {Caselli}, {Fontani},
  {Fuente}, {Krumholz}, {McKee}, \& {Stolte}}]{tan14}
{Tan}, J.~C., {Beltr{\'a}n}, M.~T., {Caselli}, P., {et~al.} 2014, in Protostars
  and Planets VI, ed. H.~{Beuther}, R.~S. {Klessen}, C.~P. {Dullemond}, \&
  T.~{Henning}, 149

\bibitem[{{Taylor} {et~al.}(2003){Taylor}, {Gibson}, {Peracaula}, {Martin},
  {Landecker}, {Brunt}, {Dewdney}, {Dougherty}, {Gray}, {Higgs}, {Kerton},
  {Knee}, {Kothes}, {Purton}, {Uyaniker}, {Wallace}, {Willis}, \&
  {Durand}}]{taylor03}
{Taylor}, A.~R., {Gibson}, S.~J., {Peracaula}, M., {et~al.} 2003, AJ, 125,
  3145, \dodoi{10.1086/375301}

\bibitem[{{Tenorio-Tagle}(1979)}]{tenorio79}
{Tenorio-Tagle}, G. 1979, A\&A, 71, 59

\bibitem[{{Thompson} {et~al.}(2004){Thompson}, {Urquhart}, \&
  {White}}]{thompson04}
{Thompson}, M.~A., {Urquhart}, J.~S., \& {White}, G.~J. 2004, A\&A, 415, 627,
  \dodoi{10.1051/0004-6361:20031681}

\bibitem[{{Urquhart} {et~al.}(2009){Urquhart}, {Hoare}, {Purcell}, {Lumsden},
  {Oudmaijer}, {Moore}, {Busfield}, {Mottram}, \& {Davies}}]{urqu09}
{Urquhart}, J.~S., {Hoare}, M.~G., {Purcell}, C.~R., {et~al.} 2009, A\&A, 501,
  539, \dodoi{10.1051/0004-6361/200912108}

\bibitem[{{Urquhart} {et~al.}(2011){Urquhart}, {Morgan}, {Figura}, {Moore},
  {Lumsden}, {Hoare}, {Oudmaijer}, {Mottram}, {Davies}, \& {Dunham}}]{urqu11}
{Urquhart}, J.~S., {Morgan}, L.~K., {Figura}, C.~C., {et~al.} 2011, MNRAS, 418,
  1689, \dodoi{10.1111/j.1365-2966.2011.19594.x}

\bibitem[{{Uyan{\i}ker} {et~al.}(2001){Uyan{\i}ker}, {F{\"u}rst}, {Reich},
  {Aschenbach}, \& {Wielebinski}}]{uyan01}
{Uyan{\i}ker}, B., {F{\"u}rst}, E., {Reich}, W., {Aschenbach}, B., \&
  {Wielebinski}, R. 2001, A\&A, 371, 675, \dodoi{10.1051/0004-6361:20010387}

\bibitem[{{Vig} {et~al.}(2014){Vig}, {Ghosh}, {Ojha}, {Verma}, \&
  {Tamura}}]{vig14}
{Vig}, S., {Ghosh}, S.~K., {Ojha}, D.~K., {Verma}, R.~P., \& {Tamura}, M. 2014,
  MNRAS, 440, 3078, \dodoi{10.1093/mnras/stu504}

\bibitem[{{Walborn} \& {Fitzpatrick}(1990)}]{walb90}
{Walborn}, N.~R., \& {Fitzpatrick}, E.~L. 1990, PASP, 102, 379,
  \dodoi{10.1086/132646}

\bibitem[{{Whitworth} {et~al.}(1994){Whitworth}, {Bhattal}, {Chapman},
  {Disney}, \& {Turner}}]{whit94}
{Whitworth}, A.~P., {Bhattal}, A.~S., {Chapman}, S.~J., {Disney}, M.~J., \&
  {Turner}, J.~A. 1994, A\&A, 290, 421

\bibitem[{{Wright} {et~al.}(2010){Wright}, {Eisenhardt}, {Mainzer}, {Ressler},
  {Cutri}, {Jarrett}, {Kirkpatrick}, {Padgett}, {McMillan}, {Skrutskie},
  {Stanford}, {Cohen}, {Walker}, {Mather}, {Leisawitz}, {Gautier}, {McLean},
  {Benford}, {Lonsdale}, {Blain}, {Mendez}, {Irace}, {Duval}, {Liu}, {Royer},
  {Heinrichsen}, {Howard}, {Shannon}, {Kendall}, {Walsh}, {Larsen}, {Cardon},
  {Schick}, {Schwalm}, {Abid}, {Fabinsky}, {Naes}, \& {Tsai}}]{wrig10}
{Wright}, E.~L., {Eisenhardt}, P. R.~M., {Mainzer}, A.~K., {et~al.} 2010, AJ,
  140, 1868, \dodoi{10.1088/0004-6256/140/6/1868}

\bibitem[{{Zavagno} {et~al.}(2010){Zavagno}, {Russeil}, {Motte}, {Anderson},
  {Deharveng}, {Rod{\'o}n}, {Bontemps}, {Abergel}, {Baluteau}, {Sauvage},
  {Andr{\'e}}, {Hill}, \& {White}}]{zava10}
{Zavagno}, A., {Russeil}, D., {Motte}, F., {et~al.} 2010, A\&A, 518, L81,
  \dodoi{10.1051/0004-6361/201014623}

\bibitem[{{Zinnecker} \& {Yorke}(2007)}]{zinn07}
{Zinnecker}, H., \& {Yorke}, H.~W. 2007, ARA\&A, 45, 481,
  \dodoi{10.1146/annurev.astro.44.051905.092549}

\end{thebibliography}


\listofchanges

\end{document}